\documentclass[conference]{IEEEtran}
\IEEEoverridecommandlockouts
\usepackage[algoruled]{algorithm2e}
\usepackage{graphicx}
\usepackage{frame}
\usepackage{enumitem}
\usepackage{subcaption}
\usepackage{float}
\usepackage{tikz}
\usepackage{caption}
 \usepackage{epstopdf}
\usepackage{listings}

\usepackage{amssymb}

\newenvironment{myalgo}[1][htb]
  {\renewcommand{\algorithmcfname}{Procedure}
   \begin{algorithm}[#1]%
  }{\end{algorithm}}

\usepackage{mathtools}

\usepackage{multirow}
\usepackage{array}
\newcolumntype{V}{>{\centering\arraybackslash} m{.25\linewidth} } 

\begin{document}

\title{Extract Method Refactoring by Successive Edge Contraction}

\author{\IEEEauthorblockN{Omkarendra Tiwari, Rushikesh K. Joshi}
\IEEEauthorblockA{\textit{Department of Computer Science \& Engineering} \\
\textit{Indian Institute of Technology Bombay}\\
Mumbai, India \\
\{omkarendra, rkj\}@cse.iitb.ac.in}}

\maketitle




\begin{abstract}
\textit{Segmentation}, a new approach based on successive edge contraction is introduced for extract method refactoring. It targets  identification of distinct functionalities implemented within a method.  Segmentation builds upon data and control dependencies among statements to extract functionalities from code by successive contraction of edges in the \textit{Structure Dependence Graph} (SDG). Three edge contractions are explored, namely \textit{structural control edge contraction}, \textit{exclusive data dependence edge contraction}, and \textit{sequential data dependence edge contraction}. The SDG is first constructed from the program, which is then collapsed into a segment graph that captures  dependence between subtasks. An intermediate representation for data and control dependencies among statements keeps the technique language independent. The approach is evaluated on four case studies, including three from the open source domain, and the findings are reported. 

\end{abstract}




\section{Introduction}
Design quality of software can be improved by restructuring. As noted by Arnold \cite{arnold1989software}, it  not  only helps  improve  comprehension  and  subsequent changes but  also helps in increasing the value of software. 
Though manual restructuring by expert designers may generate a high quality of structure, as noted by Griswold and Notkin \cite{griswold1993automated}, for  large systems, it  can take a long time with a rise in  error risk and cost. Therefore, an automation of the restructuring process becomes highly desirable.

Modern restructuring approaches focus on  design aspects  keeping in view the traditional wisdom of cohesion and coupling.  
The terms \textit{refactoring}  \cite{opdyke1992refactoring} and \textit{restructuring} \cite{mens2004survey} are commonly used to refer to such design based structural enhancements respectively in object-oriented programs and procedural contexts. Fowler \cite{fowler2009refactoring} described several kinds of code smells and the related refactorings. One of them is the \textit{long method} code smell, for which  \textit{extract method} is known to be  a suitable refactoring solution. Kim et al. \cite{kim2012field} in their investigation of refactoring practices report  that readability, maintainability, reuse,  reduction in code duplication and  bugs, and ease of  adding new features are among the primary benefits of refactoring. Murphy-Hill et al. \cite{DBLP:journals/tse/Murphy-HillPB12} analyzed  refactoring practices and observed that extract method refactoring tools are seldom used, and instead, such refactoring is carried out manually. 

In this paper, a novel graph-based approach for identifying and ranking extract method opportunities is developed and demonstrated. The approach is called \textit{segmentation}, in which, extract method opportunities form segments. The focus of the approach is on identifying and suggesting segments that are of manageable size, are cohesive and functionally sound, and are possibly reusable with little or no modification. An important aspect of refactoring activity is the role of human involvement in automation of the process, while maximizing its benefits. For example, some approaches \cite{abadi2012fine} require human support in identifying seed statements or the variable of interest, whereas, some others \cite{charalampidou2017identifying}, \cite{tsantalis2011identification} require human support in the selection of precise extract method opportunities. The first one requires human  involvement in initial phase of seeding, while the latter uses involvement in the final phase of selection. 
Our approach automatically suggests extract method opportunities, going closer to full automation, with manual intervention minimized to the final stage on need basis. 
Segmentation automatically ranks and picks up only a subset of all identified segments. In order to make the refactored code acceptable at the user end, the technique also facilitates human intervention to override a chosen segment by another suggested variant of it by enlisting the alternatives.

The source code to be refactored is represented in the form of a graph called Structure Dependence Graph (SDG). SDG is a variant of the well-known Program Dependence Graph (PDG) structure \cite{ottenstein1984program}, but it differs in the representation of control dependence among statements and blocks. Control edges present in SDG reflect the hierarchal control dependence structure of the code, as opposed to the non-hierarchal structure in PDGs. An intermediate code representation (IR) is designed primarily to bring out the  necessary structural information from the source code. It also helps in mapping the  source code to SDG. The IR is referred to as \textit{segment IR}. It helps to keep the semi-automatic refactoring technique programming-language independent. 

Segmentation transforms the SDG into a reduced version of it called \textit{segment graph}. This reduction process consists of two stages, (i) identification  of characteristic structures, and (ii) application of corresponding transformations on them using edge contraction. A vertex of a segment graph is intended to represent a cluster of statements implementing a distinct functionality. The proposed segmentation algorithm suggests extract method opportunities from segment graphs. One of the challenges in segmentation is handling of overlapping among possible opportunities under consideration. We use two affinity metrics to resolve the issue of overlapping. 

The approach is validated through the following four case studies.
Firstly, it is applied to a semi-modular version of an implementation of this  segmentation approach itself. The  experimental semi-modular source code contained about 350 lines of code in C, which included  methods of varying sizes that were synthetically created by unfolding method bodies in the original code. From this synthetically created non-modular code, our technique was more or less able to  identify the original methods.
Three open source case studies were used for validating the approach. Among them, the JUnit and the JHotDraw are two Java-based programs of Silva et al. \cite{silva2014recommending}. These are the versions that they created by inlining callee methods into the caller methods. Both the case studies have earlier been used as benchmark in the works of Silva et al. ~\cite{silva2014recommending} and Charalamidou et al. \cite{charalampidou2017identifying}. The third open source program that we used is the XData system \cite{bhangdiya2015xda}, which is used to assist in grading  answers that are SQL queries in a learning environment.

The remainder of this paper is structured as follows. Related work in recent years is reviewed in Section 2. Section 3 defines   terminologies used in developing the segmentation approach. The concept of \emph{segments} is introduced in Section 4. Section 5 demonstrates the segmentation approach through an example.
The segmentation algorithm in terms of successive edge contractions is presented in section 6. Lastly, an evaluation of the approach is presented in Section 7.

\section{Related Work}

Various restructuring approaches for Long Method code smell use techniques such as clustering \cite{xu2004program} \cite{alkhalid2010software}, control flow graphs \cite{lakhotia1998restructuring} and program slicing \cite{abadi2012fine}, \cite{kim1994restructuring}.
Lakhotia and Deprez \cite{lakhotia1998restructuring} present an approach for restructuring  by tucking sets of statements into functions. It separates two interleaved tasks into  functions. This transformation takes place in three steps called \textit{Wedge}, \textit{Split} and \textit{Fold}.

Lung and Zaman \cite{lung2004using}  present a clustering approach which exploits a variable's role in a statement to determine the type of data and control dependence associated with the statement. Conditional and iterative keywords (e.g. if, else, for, while) are also used as attributes of statements to compute similarity for clustering. Xu et al. \cite{xu2004program} present a restructuring approach using clustering based on their resemblance coefficient, and observe  desired clusters with data and control attribute ratios 5:2 and 8:3.  Alkhalid et al. \cite{alkhalid2010software} present an approach called Adaptive-KNN  which requires less computational time  as their algorithm computes similarity matrix only once. Also, they  note that results obtained for k=3 and k=5 are the same. So, the approach prefers clustering with k=3 because it further reduces the number of computations. 

Program slicing was proposed by Weiser to help in software debugging by extracting a set of statements affecting computation of values of a set of variables at a specific program point\cite{weiser1981program}. Program slicing has found its place in many different software related activities such as debugging \cite{agrawal1993debugging} \cite{xu2002dynamic}, program comprehension \cite{korel1998program}, also restructuring \cite{abadi2012fine}.
Abadi et al.  \cite{abadi2012fine} presented an approach to extract  an executable slice corresponding to a \textit{seed statement} with the help of external input. The external input includes data and control dependencies to be excluded. The extracted slice is transformed into a function around the seed statement and its closely related statements. 

Yal et al. \cite{yang2009identifying} propose an approach for identifying candidate methods for extraction based on control structures, branches, blank lines etc. The approach also suggests that the size of the candidates for extraction should be above a threshold. The candidates for extraction are ranked using coupling computation. Silva et al. \cite{silva2014recommending} propose an approach for method extraction candidates and a scoring function to rank their relevance.

The approach of Abadi et. al \cite{abadi2012fine} extracts tangled code based on information provided by user in the form of statements and variables.  The code extracted thus is likely to be specific to one functionality.
Tsantalis and Chatzigeorgiou \cite{tsantalis2011identification} present an approach based on complete computational slice that identifies the seed statements and variables without human involvement and suggests the extraction opportunity of a variable by computing union of the slices based on assignment, which are computed over all blocks for a chosen seed variable.

Charalampidou et al. \cite{charalampidou2017identifying} present a clustering approach to identify the extract method opportunities implementing a single functionality. The approach restricts the number of suggestions by ranking and grouping overlapping extract method opportunities with less than 20\% difference 
in sizes of the candidates. 
Our proposed segmentation approach is a clustering based technique as opposed to slicing based techniques. For the sake of ranking the opportunities it uses an approach similar to that of Charalampidou et al. \cite{charalampidou2017identifying}, which restricts the number of suggestions. However, in contrast to  their approach, we focus on statements that produce data instead of  output statements, when it comes to selection.
Our approach develops a novel graph based formulation to represent the structure of a program, and then clusters it into subgraphs, the likely candidates for desirable functions. The clustering uses structural metrics that are  based on structural information available in the SDG. The elements of the SDG are described in the section.

%
%

\section{Structural Representation: Segment IR and SDG}
\begin{figure*}[!h]   
\begin{minipage}{\linewidth}
\begin{minipage}{.45\linewidth}
\begin{footnotesize}
\vspace{1cm}
\begin{lstlisting}[language=C] 
A. void FiboPrime()	{
B.  int i, n, a, b, t;
C.  printf("Enter value of n (>0)");	
D.  scanf("%d",&n);
E.  a = 0;
F.  if (n ==1)
G.   printf("Fibo Term is %d",a);
H.  else	{
I.   b = 1;
J.   for (i = 3; i <=n ; i++){
K.    t = a + b;
L.    a = b;
M.    b = t;
     }
    }
N.  printf("Fibo Term is %d",b);
O.  for (i=2 ; i<=b/2;i++){
P.   if (b%i == 0)
Q.    break;
    }
R.  if (b<=1 || i <=b/2)
S.   printf("Not Prime");
T.  else
U.   printf("Prime");
   }
\end{lstlisting}
\end{footnotesize}
  \vspace{1.6cm}
\centerline{(a) Source Code}
\end{minipage}
\hspace{0.1cm}
\vline
\vline
 \hspace{0.2cm}
\begin{minipage}{.22\textwidth}
 \vspace{ 0.5cm}
\begin{small}

0. invar \\
1. input n \\
2. assign a \\
3. if n 2 \\
4. \hspace{4mm}output a \\
5. \hspace{4mm}else 3 \\
6. \hspace{6mm}assign b \\
7. \hspace{6mm}assign i \\
8.   \hspace{6mm}loop i n 4 \\
9.    \hspace{10mm}assign t a b \\
10.    \hspace{9mm}assign a b \\
11.    \hspace{9mm}assign b t \\
12.    \hspace{9mm}assign i i \\
13. output b \\
14. assign i \\
15. loop i b 2 \\
16. \hspace{5mm}if b i 1 \\
17.     \hspace{8mm}break \\
18.   \hspace{5mm}assign i i \\
19. if b i 2 \\
20. \hspace{5mm}invar \\
21. \hspace{5mm}else 1 \\
22.   \hspace{8mm}invar \\
\\
\end{small}

\centerline{(b) Segment IR}
\end{minipage}
\vline
\vline
\begin{minipage}{0.25\textwidth} 

 \includegraphics[scale = .28]{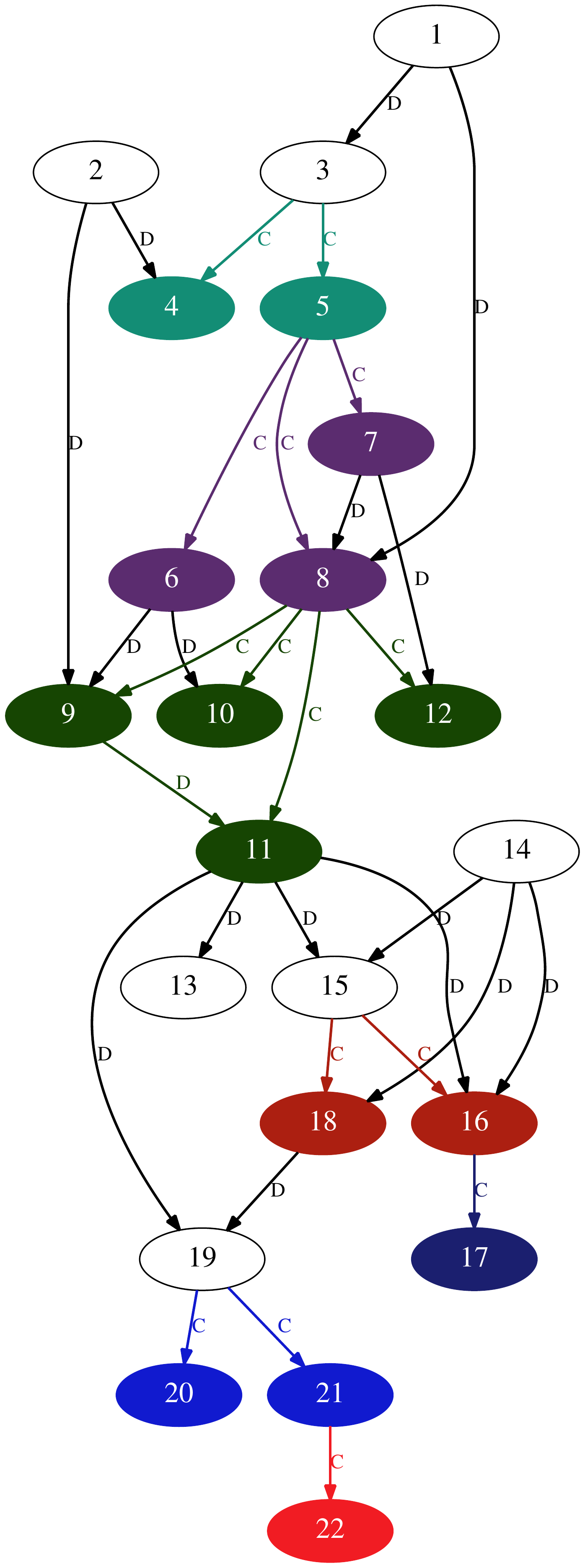}  

 \vspace{0.2cm}
 \centerline{(c) SDG}
\end{minipage}
\end{minipage}
\captionof{figure}{Intermediate Representation of Fibonacci code and corresponding SDG}
\label{fig:Fibonacci_Prime_Running_example}
\end{figure*}

This section describes representations used  in  the segmentation approach. These include an intermediate 
representation (\textit{Segment IR}) for representing programming constructs and a graph representation for data and control dependencies (\textit{SDG}). 
An example for illustrating the process of transforming source code to \textit{segment IR} and then to SDG is introduced in  Figure~\ref{fig:Fibonacci_Prime_Running_example}(a), which shows a program for finding Fibonacci Primes. Intermediate representation (described later in Section~\ref{sec:IR}) for the input source code and its corresponding \textit{structure dependence graph} (the SDG, defined later in Section \ref{sec:SDG}) are shown in Figure~\ref{fig:Fibonacci_Prime_Running_example} (b) and (c) respectively. An SDG for a segment IR is generated by mapping each IR statement to exactly one vertex, and then connecting them by edges representing two types of dependence relationships between  vertices (or statements). These relations are called \textit{structural control dependence} and \textit{data dependence}. 

\subsection{Structural control Dependence and Data Dependence } 
This section describes structural control dependence and data dependence among statements. Dependence relations among IR statements are represented by edges and their labels in the SDG.

\textit{Control statement and control block:} A statement \textit{s} is  a \textit{control  statement}  if  it is either a branching statement  or  an entry statement of a scope block enclosing one or more statements. Such structural constructs identify sets of statements that can be clustered together as a unit. Such a unit is called a \textit{control block}. Control blocks can be nested.  

The last row in Table~\ref{table:Code2SegmentIR} shows a switch-statement, where, statements under \textit{case 1} and those under \textit{default-case}  form two blocks of statements inside the outer \textit{switch} block. Other examples include \textit{if}, \textit{else} and \textit{loop} statements.


\textit{Structural control dependence:}  A statement $S_d$ is \textit{control dependent} on a control  statement $S_c$ if (i) $S_d$ is enclosed by $S_c$ and (ii) $S_c$ does not enclose any other control  statement enclosing $S_d$. It represents direct hierarchal control dependence.  

For example, in Table~\ref{table:Code2SegmentIR}, the last row contains three control  statements, namely, \textit{switch}, \textit{case} and \textit{default}, which form a unit enclosed within a \textit{switch}-statement.
This hierarchical dependence  is reflected in the corresponding IR representation, where \textit{docase}-statement(IR at line 0) encloses \textit{case}-statement  (IR at line 1) and \textit{case}-statement (IR at line 1) encloses \textit{default}-statement (\textit{case}-statement in IR at line 4).

\textit{Structural control edge:} This edge is introduced to capture the structural aspects of control in a program under refactoring. It represents the  structure of control blocks in terms of control dependencies with their internal statements. Structural control edges are different from control edges of control flow-graphs \cite{allen1970control}, in that the former show structural control dependence as opposed to behavioral control dependence in control flow graphs. Structural control edges are used in a data structure called Structure Dependence Graph (SDG).  Structural control edges result from \textit{structural control blocks} such as \textit{if-else}, \textit{for}, and \textit{do-case}. These structures are discussed in Section \ref{sec:IR}. Hereafter, the term \textit{control edge} is used to refer to structural control edges. 

\textit{Control vertex:} A control vertex represents control statements such as \textit{if-else}, \textit{elseif}, \textit{do-case}, \textit{loop} in  SDG.  A label 'C' over an edge $\langle x,y \rangle$ in  SDG represents a control edge between vertices \textit{x} and \textit{y}, where, vertex \textit{x} is a \textit{control vertex}, and vertex \textit{y} is a control dependent vertex. The same can be observed from  Table~\ref{table:Code2SegmentIR}, where, a vertex representing an if-else statement or a loop-statement in IR is attributed as a control vertex. In the table, from fourth row onwards we can observe control edges originating from the vertices corresponding to the control statements like \textit{if}, \textit{else}, \textit{do-case}, \textit{loop}. 

\textit{Data Dependence:} Statement $T$ is said to be data dependent on  statement $S$ if $T$ uses a variable defined at $S$. Data dependence between two statements is represented by a data edge with label 'D' between the corresponding vertices in SDG. Examples of the same can be observed from in Figure~\ref{fig:Fibonacci_Prime_Running_example}, where, IR statements 8 and 9 use the value of variable `n' assigned at IR 1. This dependence is represented by two data edges $\langle 1, 8 \rangle$ and $\langle 1, 9 \rangle$ in the SDG. Also, similar examples can be observed in  the first three rows of Table~\ref{table:Code2SegmentIR}.

\subsection{Segment IR}\label{sec:IR}
To support language independent refactoring, an intermediate representation called \emph{Segment IR} is defined. The transformation of input code to segment IR aims at minimizing the number of tokens used in the representation of code, at the same time preserving data and control dependence among the statements. Segment IR is the link  between the SDG and  the source code. And thus, makes the visualization of mapping the source code to graph easy and vice versa.
The IR requires only two types of information from every statement of the input code, the  operation \textit{primitive} used,  and a \textit{list} of variables and their \textit{roles} in the operation. The primitive operations considered are \textit{input}, \textit{output}, \textit{assign}, \textit{if}, \textit{else}, \textit{else-if}, \textit{do-case}, \textit{case}, \textit{loop} and  \textit{invar}.

\begin{table}
\centering
\caption{Code to Segment IR transformation}
\label{table:Code2SegmentIR}
\begin{tabular}{|p{1.2cm}|p{2.2cm}|p{1.7cm}|p{2cm}|}
\hline
\textbf{Kind} &\textbf{Example Code}  & \textbf{Equivalent  Segment IR} & \textbf{SDG Model}\\ \hline\hline

Input &scanf("\%d \%d",\&n1,\&n2);  & 0. \emph{input} n1 & \multirow{2}{*}{\includegraphics[scale=0.19]{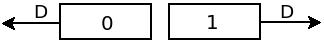}}\\
& & 1. \emph{input} n2 & \\ \hline

Output & printf("\%d \%d",n1,n2); & 0. \emph{output} n1 & \multirow{2}{*}{\includegraphics[scale=0.2]{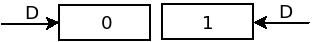}}\\
 & & 1. \emph{output} n2 & \\ \hline
 
Assignment & x = a + b & 0. \textit{assign} x a b & \multirow{2}{*}{\includegraphics[scale=.23]{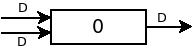}}\\
& & &  \\ \hline

If &if(x $<$ (y+10))& 0. \emph{if} x y 1& \multirow{3}{*}{\includegraphics[scale=.2]{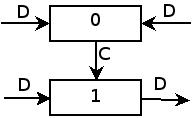}}\\ 
&\hspace{.2cm} x++; & 1. \emph{assign} x x &  \\ 
&&&\\ 
\hline
   
\hspace{-1mm}IF-Else &if (x) & 0. \emph{if} x 2& \multirow{5}{*}{\includegraphics[scale=.16]{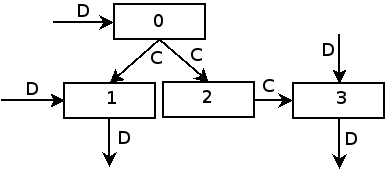}}\\
&\hspace{.2cm}y = x;& 1. \emph{assign} y x& \\
& else  & 2. \emph{else}  1 & \\
&\hspace{.2cm} x =y; & 3. \emph{assign} x y& \\ &&&\\\hline

 &if (x $>$ y) & 0. \emph{if} x y 2 & \multirow{7}{*}{\includegraphics[scale=.17]{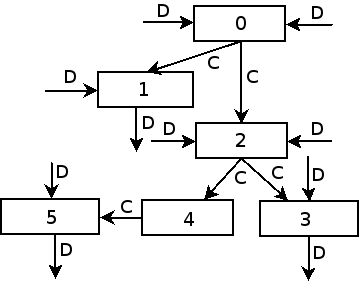}}\\
&\hspace{.2cm}v = x; & 1. \emph{assign} v x &\\
If-ElseIf-Else &else if(y $>$ z)  & 2. \emph{elseif}  y z 2 &\\
&\hspace{.2cm} v =y; & 3. \emph{assign} v y &\\
& else &  4. \emph{else} 1 & \\
& \hspace{.2cm}v = z; & 5. \emph{assign} v z&  \\
&&&\\\hline 
Loop & for (i=1;i$<=$n;i++)& 0. \textit{assign} i & \multirow{4}{*}{\includegraphics[scale=0.23]{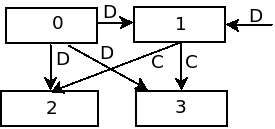}} \\ 
&\hspace{.2cm}f = f *i; & 1. \emph{loop} i n 2 &\\
& & 2. \emph{assign} f f i &\\ 
& & 3. \emph{assign} i i & \\ \hline 
&&&\\ 
Reduced Loop &for (i=1;i$<=$n;i++)& 0. \emph{assign} i& \multirow{4}{*}{\includegraphics[scale=0.2]{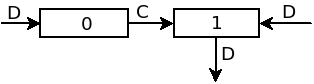}}\\
& \hspace{.2cm}f = f *i; & 1. \emph{loop} i n 1  &\\ 
 & &2. \emph{assign} f f i &  \\
 &&&\\\hline 

DoCase & switch(x)\{ & 0. \emph{docase} x 1 & \multirow{7}{*}{\includegraphics[scale=0.17]{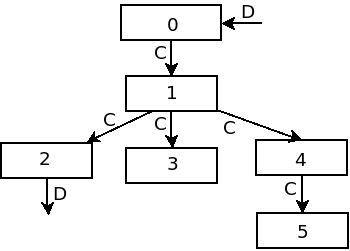}}\\
& \hspace{.2cm} case 1: & 1. \emph{case} 3 &\\
&\hspace{.4cm}s =0; & 2. \emph{assign} s &\\
&\hspace{.4cm}break; & 3. \emph{break} &\\
& \hspace{.2cm} default: & 4. \emph{case} 1 &\\
&\hspace{.4cm}printf("Default"); & 5. \emph{invar} &\\
& \} & &  \\ 

\hline
 
\end{tabular}
\end{table}

To represent conditional or iterative block statements, the segment IR  encodes the blocks by marking  the sizes of the block instead of using block markers. Actual operators are not captured, since we are interested in dependency of data and control for segmentation. 
The set of mappings for transformation of statements from input code to Segment IR is shown in Table ~\ref{table:Code2SegmentIR} through examples. The table also shows the corresponding SDG models (SDGs are explained in Section ~\ref{sec:SDG}). Each vertex in the SDG corresponds to exactly one statement in the \textit{IR} (also called segment IR). As discussed earlier, each statement in the IR is associated with an integer index/line-number. This index is used as  vertex labels in SDG.
The  primitives (keywords) included in the IR are now explained. 
 
\begin{itemize}[leftmargin=*]
 \item \emph{Input/Output:} In \textit{segment IR}, an input statement such as \textit{scanf} is represented by the keyword \emph{input}, which is  followed by the list of variables participating in the input operation. \textit{Input} is a special case of \textit{assign}. Similarly, an output statement such as \textit{printf} is represented by the keyword \emph{output}, which is followed by the list of variables participating in the output operation. If a single input or output function reads/writes  values of  multiple variables,  those variables can be listed in  multiple input/output statements, which helps the algorithm exercise finer control in identifying segments. For example, in the table, two inputs in \textit{scanf} statement are split in two statements in the \textit{segment IR}. The same can be observed for \textit{printf} statement in the table. However, such separation of variables is not always required, as in the case of file input/output, in which a file-pointer can not be separated from program variables used for file data. Such a group of variables can be listed together in a single IR statement. The algorithm does not split them. For example, in 
 Figure~\ref{fig:Fibonacci_Prime_Running_example} (a) and (b), the \textit{scanf} statement at index D is represented in IR at line 1, and the \textit{printf} statements at lines G and N are represented in IR at lines 4 and 13. It can be noted here that statement C does not use any variable. Hence, it is represented in IR by an \textit{invar}-statement.
 
 \item \emph{Assign:} An assignment statement is represented by  keyword \emph{assign}, which is  followed by a list of variables participating in the assignment operation. The first variable in the list represents the l-value variable that is being modified and  rest of the variables in the list are variables  accessed  as r-values.  The table shows an example of this case with two inputs and a result.
 In Figure~\ref{fig:Fibonacci_Prime_Running_example}, this representation can be observed for statements K, L and M of source code, where they are represented in the IR on lines  9, 10 and 11 respectively.

\item \emph{If, Else and ElseIf:} The \textit{if} follows a list of variables and an integer. The list contains all the variables participating in decision regarding execution of the block. An integer defines the block scope as the count of immediately following  statements dependent on it (block length).
Block lengths are also noted for the \textit{else}  and the \textit{elseif} blocks. Separate examples are provided in  the table for the three statements. 
Figure~\ref{fig:Fibonacci_Prime_Running_example} shows the if-else statement on lines F and H represented in IR on lines 3 and 5.  Also, if-else on lines R and T is  represented in IR on lines 19 and 21.

 \item \emph{Loop:}  The iterative statements can be represented in one of the two forms. In the first form, three components, namely initialization, condition checking, and increment/decrement are represented separately. On the other hand, the \textit{reduced loop} representation does not consider the loop step operation. 
 The latter may possibly reduce  the number of edge contractions in the algorithm resulting  in coarser segments. Both loop representations use the keyword \emph{loop} followed a list of variables and an integer. The list contains all the variables participating in the decision making and the integer defines the scope, that is the number of dependent statements. The examples for both are shown in the table. The reduced loop representation is not used in this paper.
 Figure~\ref{fig:Fibonacci_Prime_Running_example} illustrates transformation of a loop statement on line J into three IR statements 7, 8 and 12. The same transformation can be seen for another for-loop statement O, which is transformed into IR statements 14, 15 and 18.
  
  \item \emph{DoCase:}  The \emph{docase} keyword is followed by a variable. The \textit{case} statements and the \textit{default} statement are represented by the keyword \emph{case}. It is noted that no special keyword is required for \textit{default} case, since we are only interested in SDG. An example is shown in the table with one case and a default block.
 
 \item  \emph{Invar:} It is a statement that does not use a variable.
 For example, \textit{print} statement which uses constants is an {\em Invar} statement. A few instances of such a statement are source code statements C, S and U, which are represented as IR statements 0, 20 and 22 respectively as shown in Figure~\ref{fig:Fibonacci_Prime_Running_example}. 
  
\end{itemize}

  Table~\ref{table:IRQueryFunction} lists with examples a set of functions to query a variable's role over a set of statements. In the next section, the concept of SDG is developed in terms of these functions.

\subsection{Functions for querying IR}
  Table~\ref{table:IRQueryFunction} lists with examples a set of functions to query a variable's role over a set of statements. In the next section, the concept of SDG is developed in terms of these functions.

\begin{table}[h]
\caption{Functions for querying IR}
\label{table:IRQueryFunction}
\begin{tabular}{|p{5cm}|p{3cm}|}
\hline
 \textbf{Functions} \& \textbf{Description} & \textbf{Instance from Figure~\ref{fig:Fibonacci_Prime_Running_example} (b)}\\ \hline
 
 \textit{DefinedAt (id)}:  
 It returns the set of variables defined (i.e., assigned) in IR statement at index \textit{id}. & \textit{DefinedAt (1) $\rightarrow$ \{n\}}
 
 \textit{DefinedAt (9) $\rightarrow$ \{t\}}\\ \hline
 
 \textit{UsedAt (id)}:  
 It returns the set of variables of which the values are used in IR statement at index \textit{id}. & \textit{UsedAt (9) $\rightarrow$ \{a,b\}}\\  \hline

 \textit{LastDefined (var, id)}:  
 It returns the id of the statement where last occurrence of definition (i.e., value assignment) for \textit{var} is found prior to  statement at index \textit{id}. That is the last location in range $0..id-1$ where \textit{var} is defined. & \textit{LastDefined (b, 13) $\rightarrow$ 11} 
 
 \textit{LastDefined (b, 11) $\rightarrow$ 6} 
 \\ \hline

 \textit{IsControlBlock (id)}: 
 It returns \textit{True} if the primitive in statement at \textit{id} is a control block such as \textit{if-else}, \textit{do-case}, \textit{for}, \textit{while}. All control block primitives  include the length of the IR-block as given in Table~\ref{table:Code2SegmentIR}. & \textit{IsControlBlock (5) $\rightarrow$ True}
 
 \textit{IsControlBlock (6) $\rightarrow$ False}
 \\ \hline
 
 \textit{GetCtrlBlocks (startId, endId)}:  
 It returns the set of ids of all control  statements appearing in the index range $startId+1..endId-1$. & \textit{GetCtrlBlocks (3, 17) $\rightarrow$ \{5, 8, 15, 16 \}} \\ \hline
 
 \textit{GetLength (id)}:  
 It returns the length of the given control  statement. It is found at the end of the control  statement located at \textit{id}. & \textit{GetLength (8) $\rightarrow$ 4}
 
 \textit{GetLength (5) $\rightarrow$ 3}
 \\ \hline
 \textit{GetLengthSum (id1, id2)}: $$ \sum_{\forall b \in GetCtrlBlocks (id1, id2)} {getLength (b)}$$ & \textit{GetLengthSum (2, 14) $\rightarrow$ 9} 
 \textit{GetLengthSum (3, 14) $\rightarrow$ 7} 
  \\ \hline
 \textit{IsControlParent (pid, cid)}: 
   It returns \textit{True} only if statement at index \textit{cid} is directly inside the  control block defined at index \textit{pid}.
    For a given pair of IR indices  $cid>pid$, this function can be represented by the  following condition: 
 
   $((cid-pid)-GetLengthSum (pid,cid))\leq GetLength(pid)$  
   &  \textit{IsControlParent (8, 12)$\rightarrow$ True}
   
   \textit{IsControlParent (5, 8) $\rightarrow$ True}

   \textit{IsControlParent (5, 9) $\rightarrow$ False}  
  \\ \hline

\end{tabular}
\end{table}

\subsection{Structure Dependence Graph (SDG)} \label{sec:SDG}

An SDG is a directed graph, which represents a mapping of an IR specification. IR statements are mapped to vertices, and dependence among statements into control dependence or data dependence edges are mapped to labeled edges connecting their corresponding vertices. Vertices use labels of their corresponding IR statements. Edge labels \textit{`C'} and \textit{`D'} represent control and data dependence respectively. Let a directed graph  $G = (V,E)$  be the SDG corresponding to an IR specification $S$. Let $E=E_d \cup E_c$, where $E_d$ and $E_c$ are sets of edges with label `D' and label `C' respectively. Let $ID$ be the set of indices of statements in $S$.
 The mapping of $S$ to $G$ is given by the
following mapping rules defined in terms of functions listed in Table \ref{table:IRQueryFunction}. 

\subsubsection{Rules for Mapping IR to SDG:}

\begin{enumerate}
 \item  Every statement in the IR is mapped to a unique vertex.
 
 $ u \in ID \implies u \in V$.
 
 \item For every statement that uses of a variable in IR, an edge labeled `D' is inserted in G to the vertex corresponding to this statement from the vertex corresponding to its value source.
 
 $ u,v \in ID$, $\exists var \in (UsedAt(v) \cap  LastDefined(var,v) = u) \implies \exists \langle u, v \rangle \in E_{d} $

 \item For every statement that is inside a control block, an edge labeled `C' is inserted in G to the vertex corresponding to this statement from the vertex corresponding to its parent control statement.

 $ u,v \in ID$, $IsControlParent(u,v) \implies \exists \langle u, v \rangle \in E_{c} $  

\end{enumerate}

\subsubsection{Example:}

Figure~\ref{fig:Fibonacci_Prime_Running_example} (c) shows the SDG corresponding to  segment IR Figure~\ref{fig:Fibonacci_Prime_Running_example} (b) of the Fibonacci prime program Figure~\ref{fig:Fibonacci_Prime_Running_example} (a).  IR statement at index 4 and 5 are direct control dependent on IR statement at index 3 (if-statement), whereas IR statements 6, 7, and 8 are direct control dependent on IR statement 5 (else-statement). This direct control dependence is captured by a directed control edges (shown as edges with label `C') $ \langle 3, 4 \rangle, \langle 3, 5 \rangle, \langle 5, 6 \rangle, \langle 5, 7 \rangle, \langle 5, 8 \rangle $. In the figure, for the sake of clarity, all vertices sharing the same control parent have been shown in same color. For example, vertex 3 is control parent of 4 and 5, and vertex 9 is control parent of vertices 9-12.
Similarly, data edges $\langle 1, 3 \rangle $ and $\langle 1, 8 \rangle $ reflect the use of variable \textit{n} at index 3 and 8, and its value assignment at index 1 in the IR. Also, an edge $\langle 11, 13 \rangle$ reflects the consumption  at line 13, of value generated at statement 11. This edge corresponds to variable \textit{b}.

\subsection{Properties of SDG:}
  This subsection describes properties of a SDG, which are used in the refactoring process using segmentation approach. 
\subsubsection{Source vertex:} \label{Sec:Source}

A \textit{source} vertex is a vertex with no incoming data edges and at least one outgoing data edge. There can be more than one \textit{source} vertices in a SDG. A source vertex in SDG captures an input statement or definition of a variable with constant in the program.

\subsubsection{Chains:} \label{Sec:Chain}

\begin{figure}
\begin{tabular}{ccccc}
\includegraphics[scale = .23]{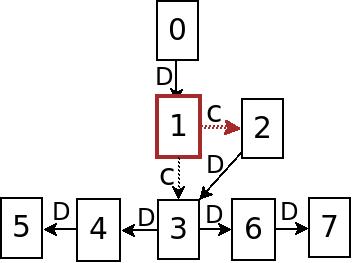}&
\includegraphics[scale = .23]{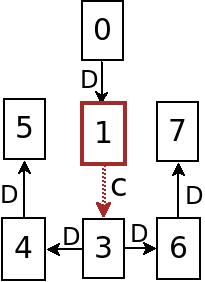} &
\includegraphics[scale = .23]{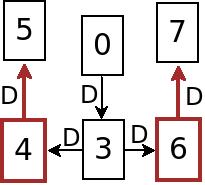} &
\includegraphics[scale = .23]{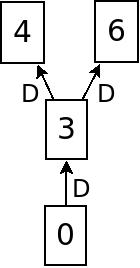}\\
(a) & (b) & (c) & (d)
\end{tabular}
\caption{Example graph for chains and Edge Contraction}
\label{fig:ChainExample}
\end{figure}

They define one type of collapsible structure in our model. 
It is a directed path  $u_{0}\rightarrow u_{1}\rightarrow...u_{k-1}\rightarrow u_{k}$ consisting of vertices and data edges in the SDG such that each vertex except $u_0$ has exactly one predecessor, and each vertex except  $u_k$ has exactly one successor. The vertices $u_0$, $u_k$ are called \textit{head}, \textit{tail}  of the \textit{chain} respectively. Such a structure captures sequential data dependence in the program and may correspond to a subtask. Length of the \textit{chain} is measured by the total number of edges present in it. In Figure ~\ref{fig:ChainExample}(a), paths\{$2\rightarrow3$\}, \{$4\rightarrow5$\}, \{$6\rightarrow7$\} are chains, while paths \{$0\rightarrow2\rightarrow3$\}, \{$3\rightarrow4\rightarrow5$\} are not.

\subsubsection { Edge Contraction, Seg Id:} \label{Sec:SegId}

Edge contraction is an operation which involves removal of an edge and merging of its two end vertices. As a result a new vertex is created. All the edges incident on either of the end vertices of the contracted edge are mapped to a single new vertex. After edge contraction, the resulting vertex is assigned a new label called as \textit{segment id} (or seg id). 
For an edge $\langle u,v\rangle \in E_{G}$, which is being contracted, the seg id is identified as follows:

\begin{enumerate}
 \item $u$ is seg id if $\langle u,v\rangle$ is a control edge, else
\item  $v$ is seg id vertex if $u$ is a source vertex, else 
  \item  $u$ is seg id if $\langle u,v \rangle$ is a part of \textit{chain}
\end{enumerate}

 An example  transformation of an SDG by a sequence of edge contractions (of colored edges) is shown through  Figures~\ref{fig:ChainExample}  (a), (b), (c) and (d). Colored edges show the edges to be contracted, and  colored vertices show the \textit{seg ids} for the resulting vertices.

\subsubsection{Control depth, Control region:} \label{sec:control_region}

The \textit{control depth} of a vertex $v$ from a source vertex $u$ is the number of \textit{control edges} present in the path from $u$ to $v$. For example, in Figure ~\ref{fig:Fibonacci_Prime_Running_example}, control depth of vertex 6 and 7 from the source vertex 1 is two. 
The \textit{control region} of a vertex reflects the control block to which the  vertex belongs. Control region of a vertex $v$ is the vertex-id of it's control parent.
It is set to -1, if it is not control dependent on any other vertex, indicating it is not part of any control block, i.e., it resides in the outermost control region.
 In other words, if a vertex $v$ has an incoming control edge (edge with label `c') from some vertex $u$ then its control region is `u', and  it is -1 otherwise. Vertices in same color in Figure~\ref{fig:Fibonacci_Prime_Running_example} share the same control region. For example, vertices 4 and 5 are in same  region 3, whereas 9, 10, 11, and 12 are in region 8. 

\begin{figure*}[h]   
\begin{minipage}{\linewidth}
\begin{minipage}{.35\linewidth}
\begin{lstlisting}[language=C] [basicstyle=\footnotesize]
int ctrlRegionDemo()
{
  int n1,n2=0;
  scanf("%d",&n1);
  if (n2!=1)
  {
    scanf("%d",&n2);
    printf("%d",n1%n2);
  }
}
\end{lstlisting}
 \vspace{.1cm}
\centerline{(a) Source Code}
\end{minipage}
\vline
\vline
\vline
 \hspace{0.5cm}
\begin{minipage}{.25\textwidth}
\vspace{9 mm}
0. assign n2 \\
1. input n1 \\
2. if n2 2 \\
3. \hspace{2mm} input n2 \\
4. \hspace{2mm} output n1 n2 \\
\\\\\\\\
\centerline{(b) Segment IR}
\end{minipage}
\vline
\vline
\begin{minipage}{0.25\textwidth} 
\vspace{-0.9cm}

 \includegraphics[scale=0.35]{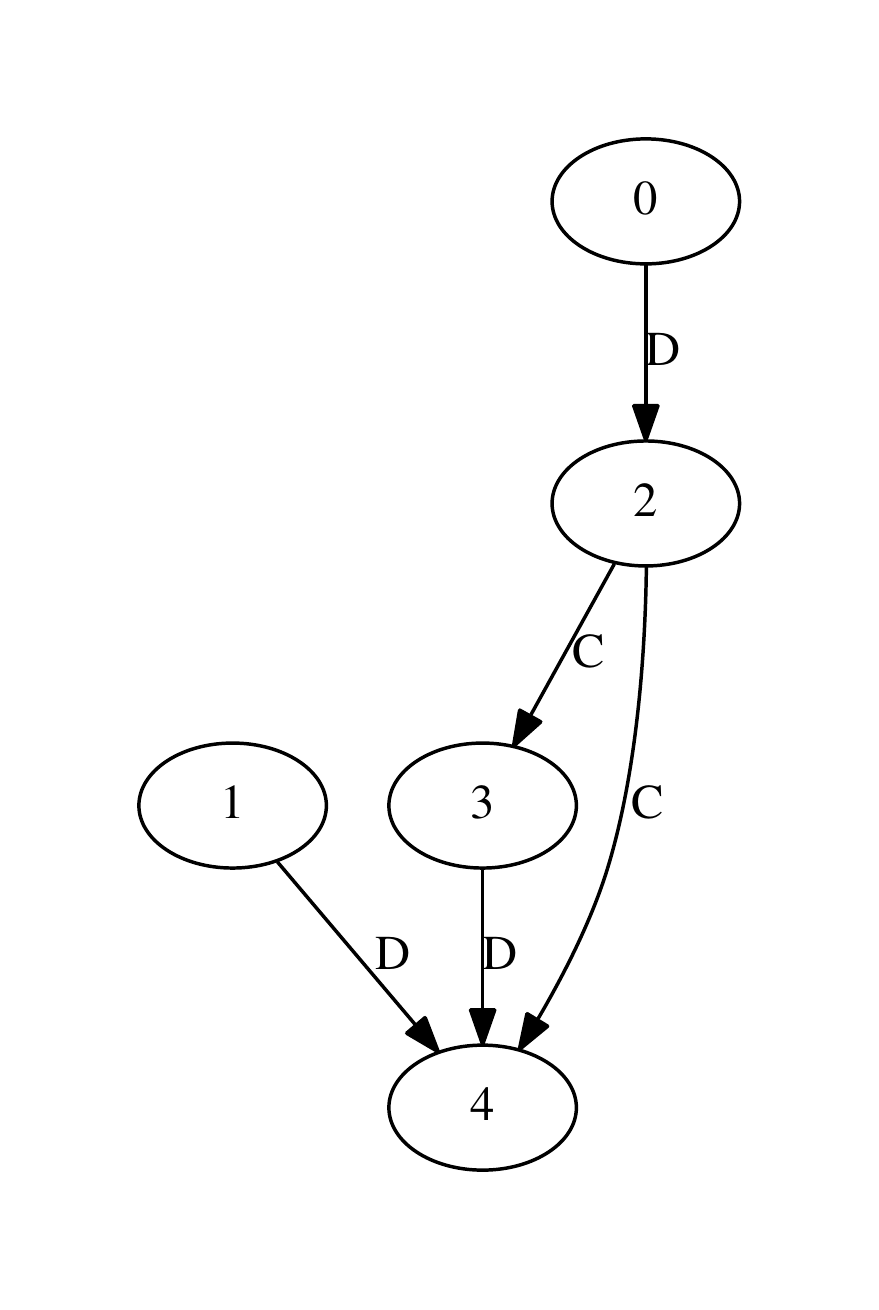}
  \vspace{-0.7cm}
 \centerline{(c) SDG}
\end{minipage}
\end{minipage}
\captionof{figure}{Control region example}
\label{fig:control_region_example}
\end{figure*}

The concept of control region permits detection of  data edge head vertices that can be merged into tail vertices without altering program behavior.  Edge contraction is not permitted when head and tail vertices of a data edge do not belong to the same control region.
In Figure~\ref{fig:control_region_example},
a source code, it's segment IR and SDG are shown in (a), (b) and (c) respectively. Statement ids in (b) correspond to vertex ids in (c). As shown in (c), vertices 3 and 4 are in the same control region 2, whereas, vertices 0, 1 and 2 are under control region  -1, they being not control dependent on another vertex. Vertex 4 has an incoming exclusive source data edge from vertex 1, indicating a closed data association between the two. But vertex 1 can't be merged into vertex 4, without moving the corresponding statement into the if-block denoted by control region 2, since it would change the program's behavior.  On the contrary, vertex 3 can be merged into vertex 4 as they are in the same control region  2, and hence
 the exclusive data edge to vertex 4 from vertex 3 can be contracted.

\section{Segments}\label {Sec:SegSlicing}

The refactoring process requires structural enhancement, which can be done by binding closely related statements into separate units and thus lowering their interdependency. Now, we discuss the approach to  extract such modular units as \textit{segments} from the SDG graph, which in turn is extracted from the IR corresponding to input code. Now, we introduce the notions of \textit{control independence} and \textit{data independence} as guidelines to identify subgraphs of SDG which are extractable as new methods.

\subsection{Control independence}

\textit{Primary and Secondary Control Vertices:} A control vertex is either a primary or a secondary control vertex. A vertex representing one of \textit{If, Loop, or DoCase} statements is called \textit{primary control vertex}.  A  vertex representing one of \textit{Else, Elseif, Case, break, or continue} statements is called a \textit{secondary control vertex}, since such control statements are control dependent on a \textit{primary control vertex}. 

Before discussing the definition of control independence the following properties of SDG are noted:

\begin{enumerate}
 \item A secondary control vertex can be direct predecessor of a control vertex only if they share at least one primary control vertex as a predecessor.
 \item A secondary control vertex can have exactly one primary control vertex as direct predecessor.
\end{enumerate}



Let G = $(V_G,E_G)$ be the SDG, and C = $(V_C,E_C)$ be a weakly connected subgraph (i.e., connected not withholding the directions) of G. Graph C is \textit{control independent} if, (i) no vertex in $V_{G-C}$ has an incident \textit{control} edge from a vertex in $V_C$ and (ii) no vertex except a primary control vertex in $  V_{C}$ has an incident control edge from a vertex in $V_{G-C}$. 
For example, in Figure~\ref{fig:Fibonacci_Prime_Running_example}, weakly connected subgraph C = (V,E), where V = \{21,22\} and E = {$\langle21,22\rangle$}, is not control independent as primary control vertex for vertex 21 is not present in C.


%

\subsection{Data independence:}

A weakly connected subgraph of SDG is data independent if every data edge that is outgoing from  or incoming into the subgraph connects two vertices in different control regions. In other words, in a data independent unit,  all its data receivers and their suppliers stay together if their control region is the same. 

%

Let G=$(V_G,E_G)$ be the SDG, D=$(V_D, E_D)$ be a weakly connected subgraph of G. Graph D is \textit{data independent} if for any  $u \in V_D, \nexists w \in V_{(G-D)}$ such that data edges  $(\langle u, w \rangle \in E_G $ or $ \langle w, u \rangle  \in E_G)$ and $u$ and $w$ have the same control region.

 For example, in Figure~\ref{fig:Fibonacci_Prime_Running_example}, weakly connected subgraph D = (V, E), V = \{$8, 9, 10, 11, 12, 13$\}, E= \{$\langle8, 9\rangle, \langle8, 10\rangle,\langle8, 11\rangle, \langle8, 12\rangle, \langle9, 11\rangle,\langle11, 13\rangle$\} is data independent.
 
\subsection{Segments}

 A \textit{segment} is a subgraph S of SDG G, which is both control independent and data independent in G. Vertices present in a segment can be classified in the following three categories. 

%


 (i) \textit{Producer} Often a segment contains statements providing data to other statements in the segment to accomplish their tasks. Such statements are called \textit{producer} statements. In a segment, such statements are represented by vertices which share data edges with other vertices. 
 (ii) \textit{Consumer} A set of statements involved in computations of intermediate and final results are \textit{consumer} statements. These vertices  are data dependent directly or indirectly  on \textit{producer} vertices. 
 (iii) \textit{Relays} A set of statements communicating computed results with other segments are called \textit{relays}. Thus, these are also \textit{producer} vertices.

It can be observed that \textit{relay} statements in a segment contain the final result of the segment computation. Thus, a relay statement can also be a  \textit{consumer} statement.  Desirable and undesirable segments can be characterized in terms of producers, consumers, and relays. For example, segments having more than one relay statements may not be desirable.


\section{Overview of the Segmentation Approach}\label{sec:Example}
While the detailed segmentation algorithm is developed in the next section,  
this section provides an overview of the process of refactoring using the  segmentation approach through an example of   \textit{Fibonacci Primes} as shown in Figure~\ref{fig:Fibonacci_Prime_Running_example}. 
The approach builds around \textit{control blocks} to identify functionalities from the SDG.  

IR for the input source code and its corresponding \textit{structure dependence graph} are shown in Figure~\ref{fig:Fibonacci_Prime_Running_example} (a) and (b) respectively. Figure~\ref{fig:FiboPrime} shows the input code and  the refactored code after applying segmentation.
Transformation of  this SDG to a graph called \textit{segment graph} is achieved by a process of \textit{successive edge contraction} as illustrated below.

\begin{figure}
 \begin{tabular}{p{2.7cm}p{2.2cm}p{3cm}}
  \includegraphics[scale = .2]{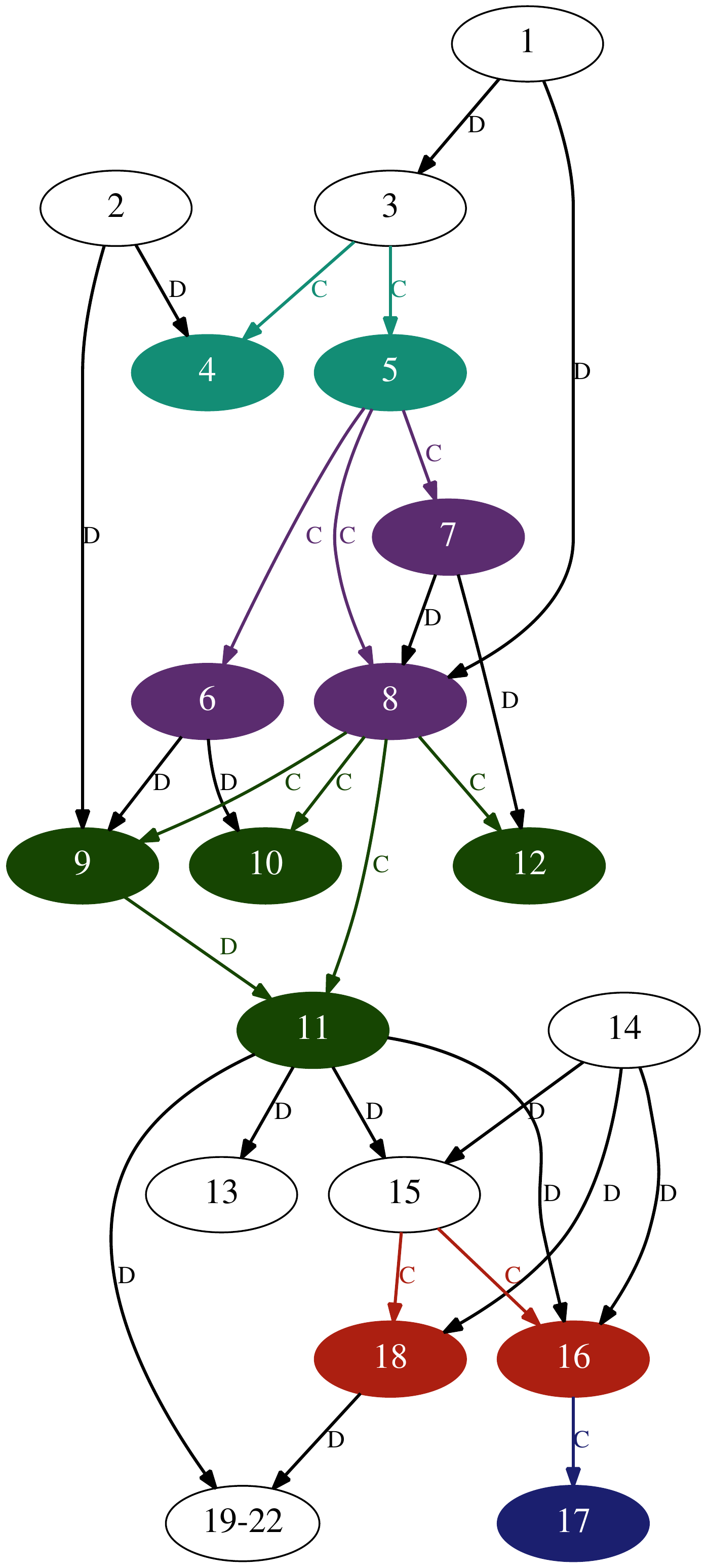} &  \includegraphics[scale = .2]{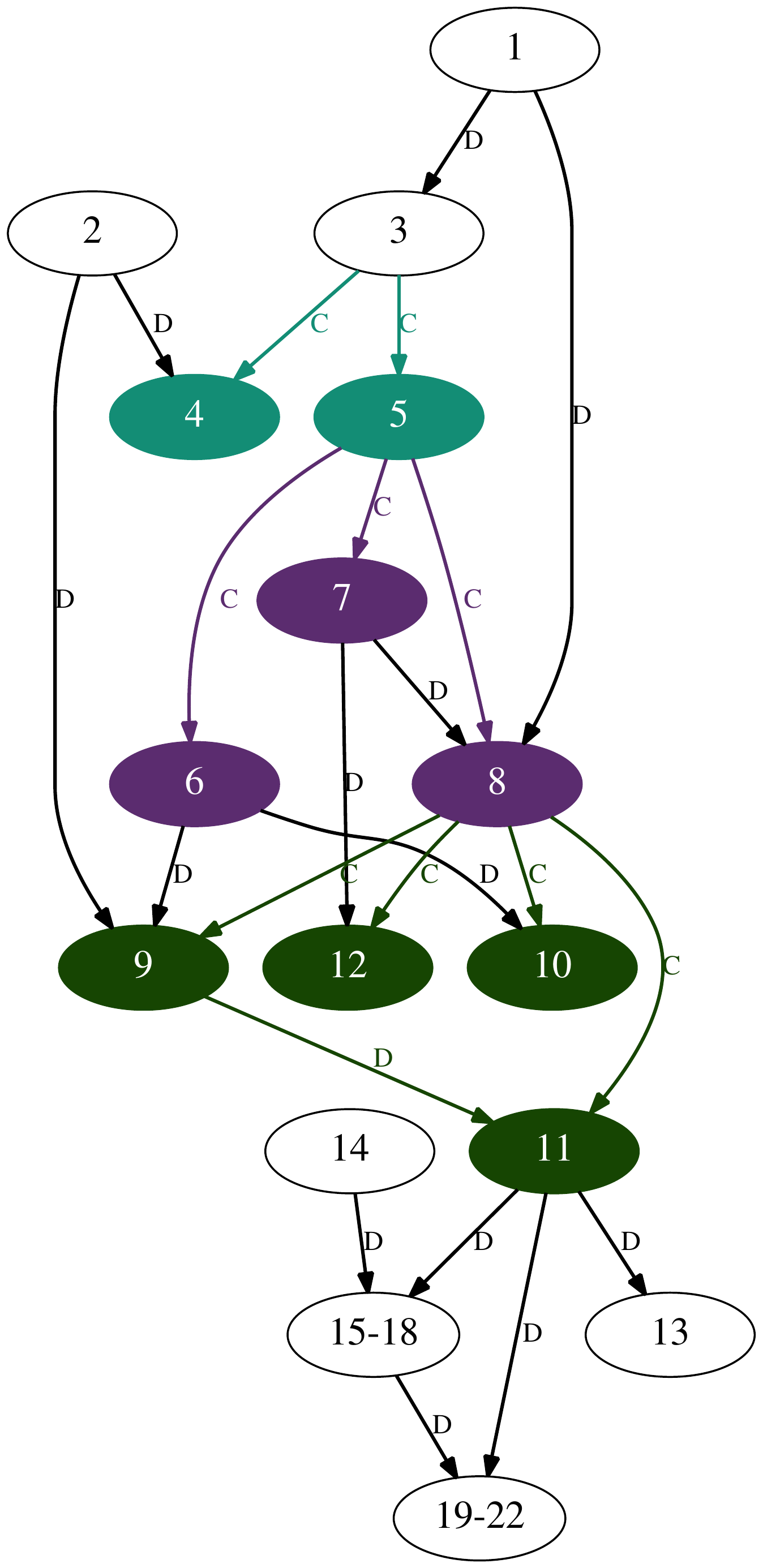} & \includegraphics[scale = .2]{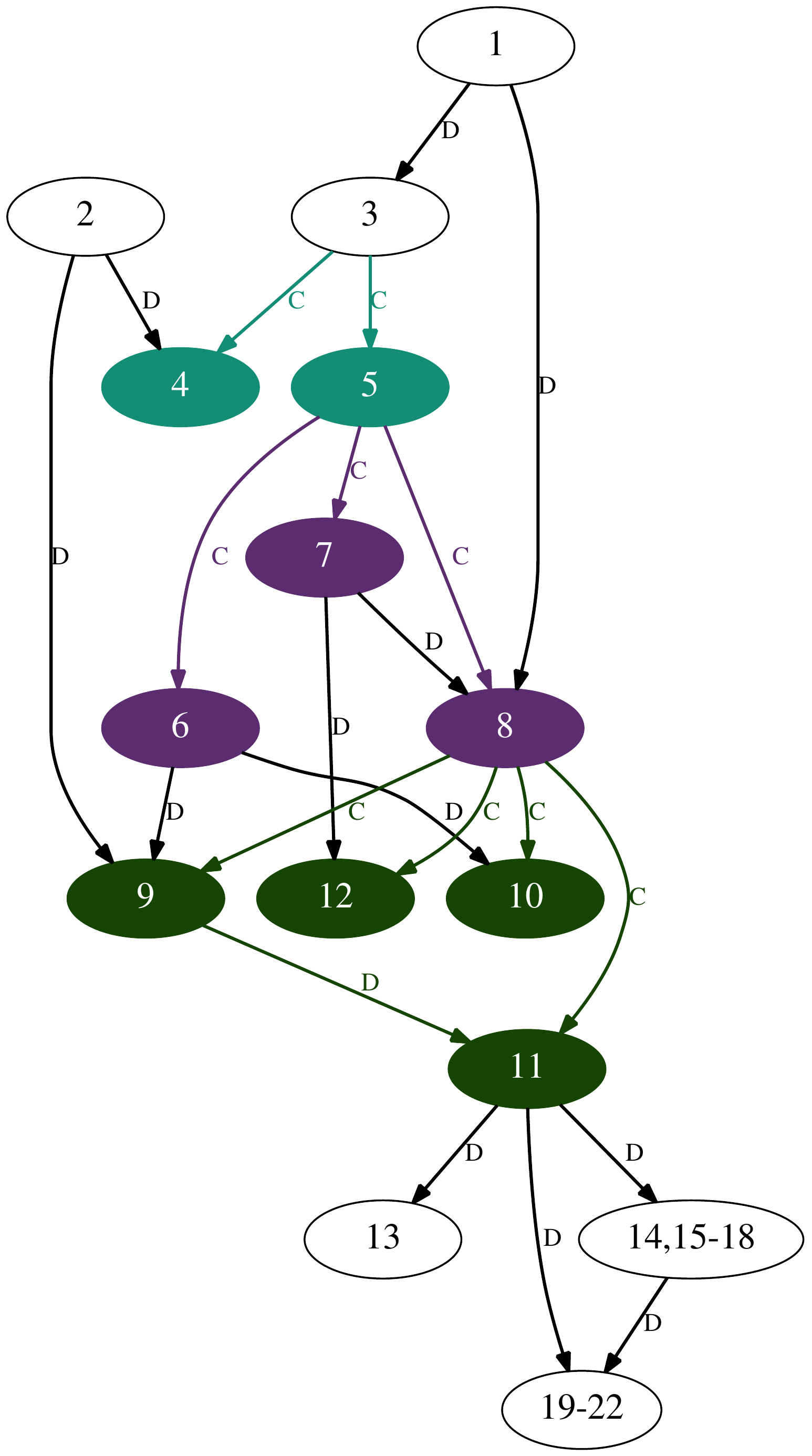}\\ 
  (a) Contracting Block at `19' & (b) Contracting Block at `15' & (c) Exclusive Source Contraction at `15' \\ 
  \includegraphics[scale = .25]{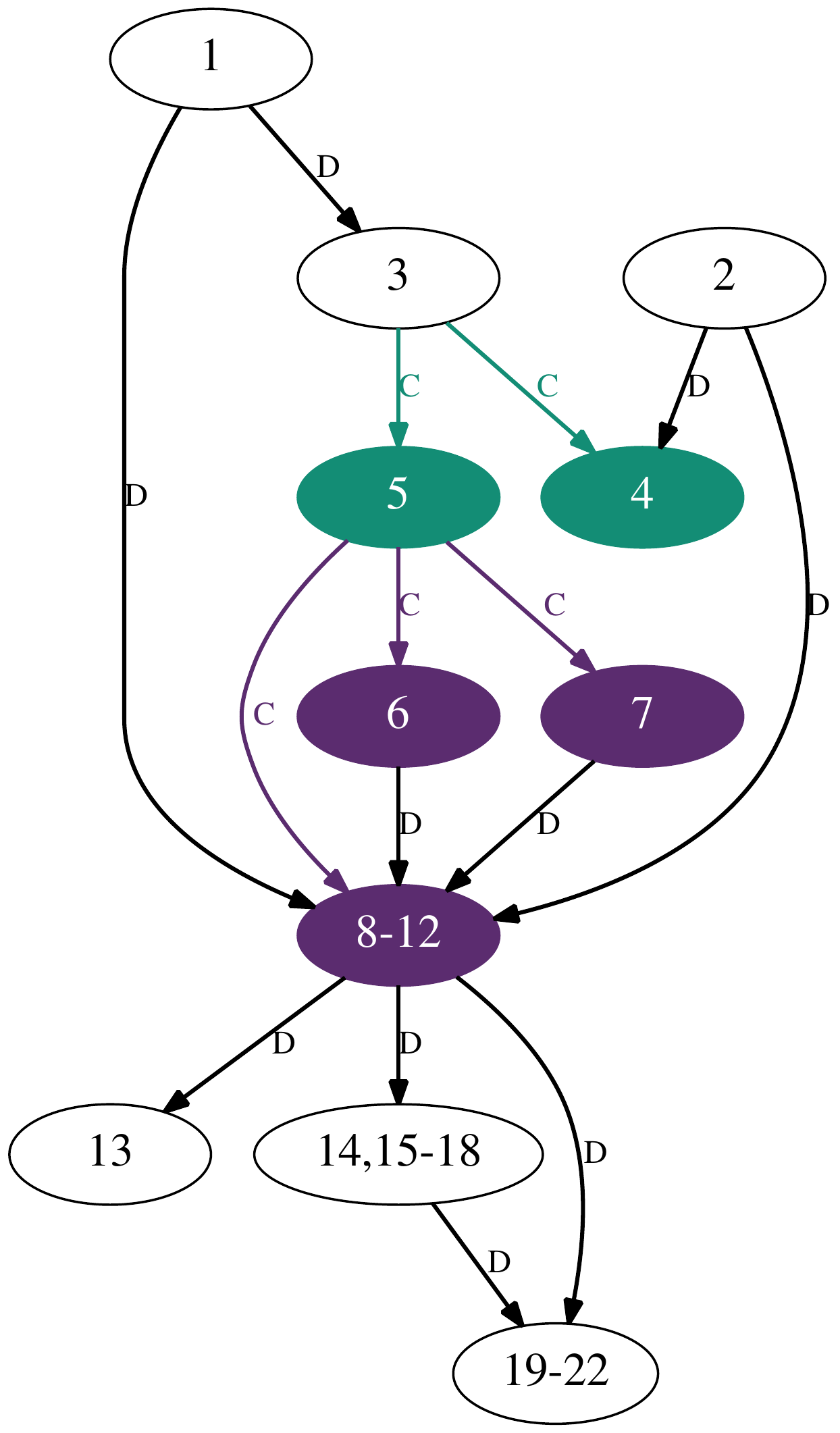} & \includegraphics[scale = .25]{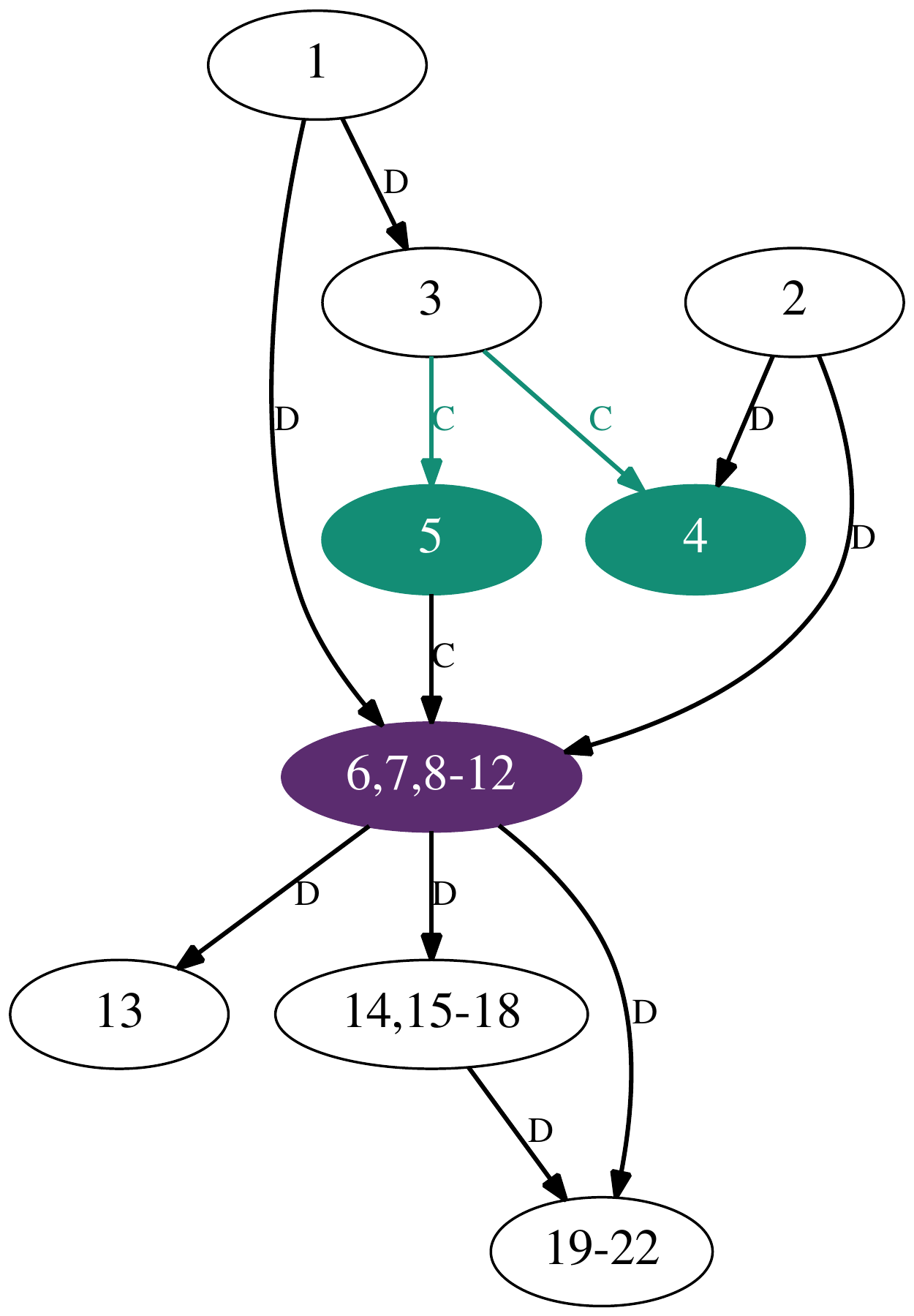}& 
  \includegraphics[scale = .3]{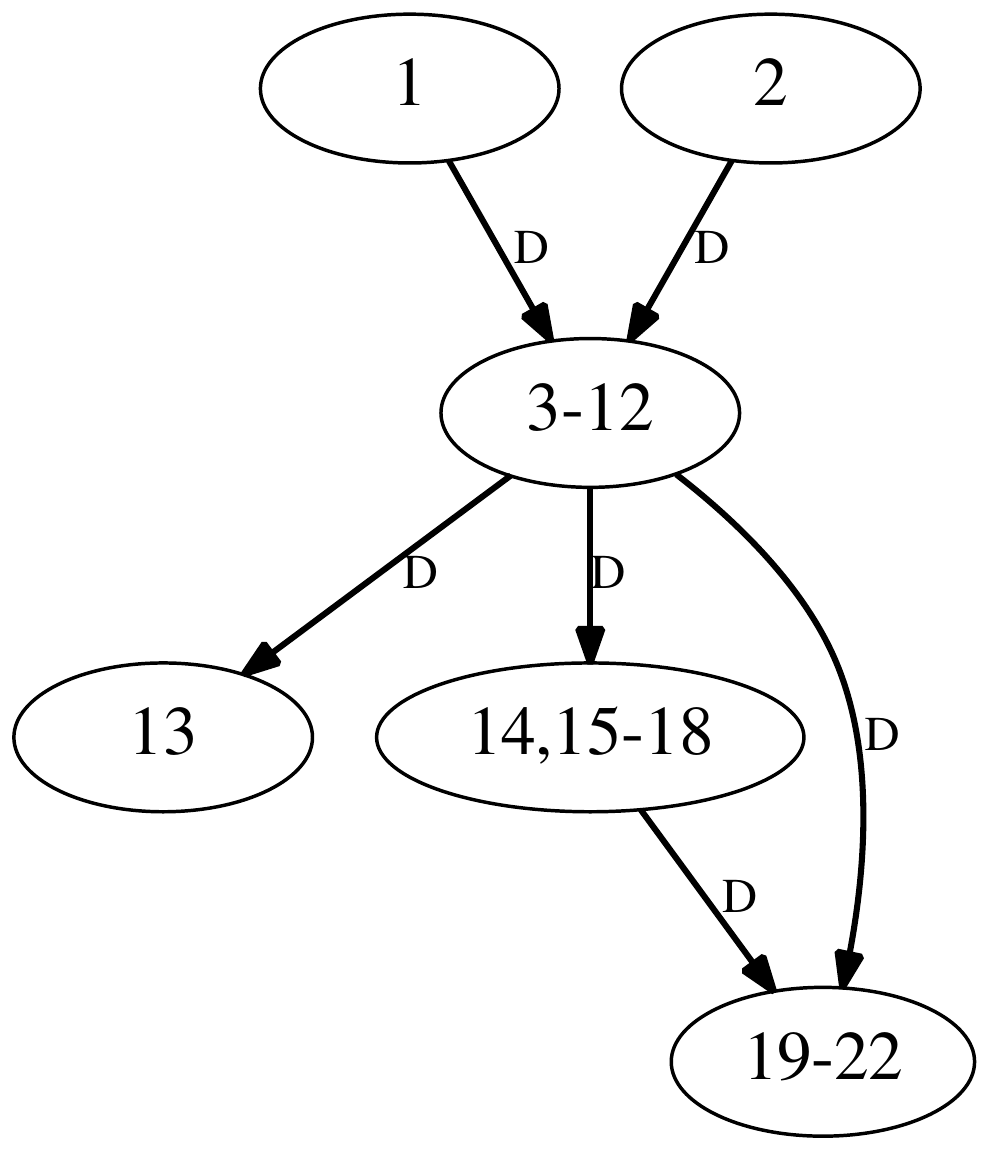}\\
  (d) Contracting Block at `8' &(e) Exclusive Source Contraction at `8' & (f) Contracting Block at '5' and `3'\\ 
    \includegraphics[scale = .31]{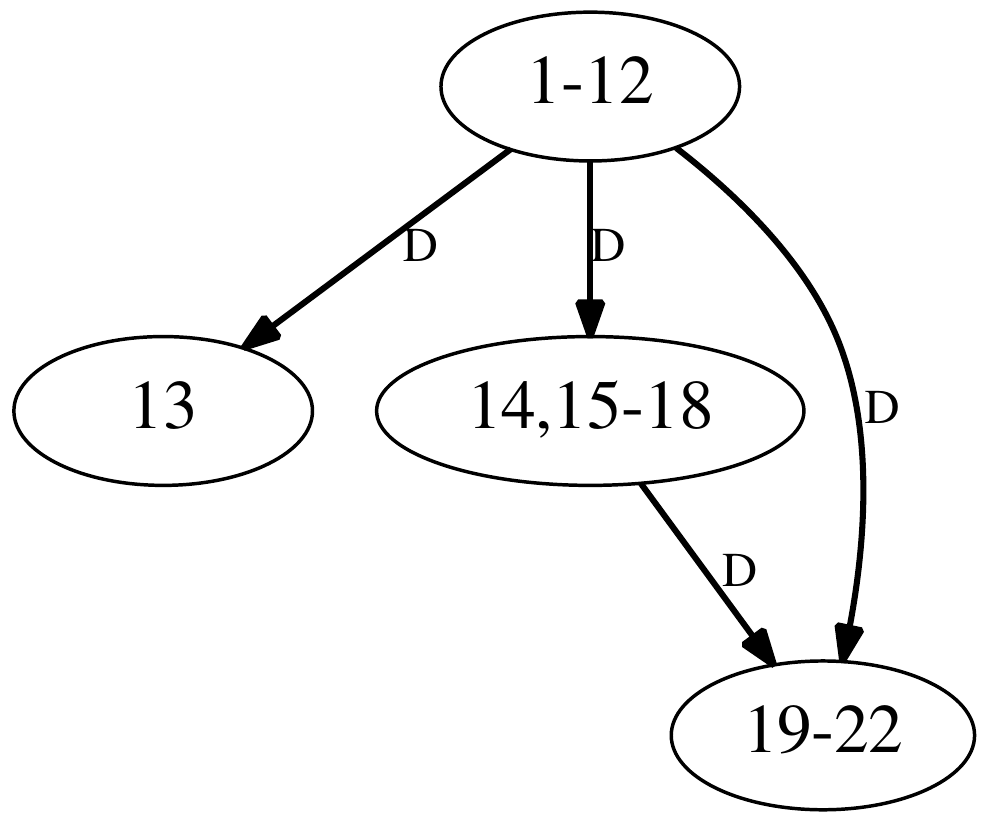}& \includegraphics[scale = .3]{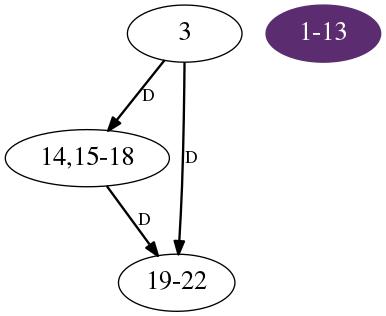} &
  \\
  (g)  Exclusive Source Contraction and extraction of `3' & (h) Chain contraction and extraction of `3' &\\ 
   
  
 \end{tabular}
 \caption{Structure Based Refactoring using Segmentation}
\label{fig:contractingBlocks}
\end{figure}

\textit{Identifying control blocks:} The SDG shown in Figure~\ref{fig:Fibonacci_Prime_Running_example} contains vertices and edges in multiple colors. In the figure, a set of statements represented by vertices of a particular color (non-white) represent a single control block. It can be seen that all statements in a control block are dependent on a common control vertex. For example, vertices \{4,5\} and \{6,7,8\} form two separate control blocks that are control dependent on control vertices 3 (if-block)  and 5 (else-block) respectively. 


\textit{Identifying extract method opportunity:} After the SDG is generated, computation intensive control blocks are identified. In the SDG, for our purpose, a \textit{computation unit}  is identified by data dependence, which is a transfer of a data value from an \textit{assignment} to its \textit{use}. For identifying computation intensive units,  the approach investigates control blocks in bottom up order. Each block with computation units is a candidate representing a functionality. However,  not all such functionalities can be suggested for extraction as a separate method. So, each control block is measured in terms of a computational strength and its affinity with its parent control block. For this purpose two quality metrics named as \textit{lack of computational strength} (LoCS) and \textit{parent affinity} (PA) are defined. These are described in Sections~\ref{Sec:LoCS} and \ref{sec:PA}. 

The LoCS value for a selected control block must be below a threshold value (in our case, 0.41) for it to be considered for extraction as a separate method. Once the block qualifies the LoCS threshold, it is further considered for merger with its parent block. The PA metric investigates if the control block implements a functionality which is closely associated with parent block, to decide whether the newly identified candidate for extraction should be merged with the parent block.

If the selected control block qualifies the PA threshold, it is collapsed into a single parent vertex by contracting all the control edges present in the block. Further, the newly collapsed block is investigated for a possible merger with statements that supply data to the newly contracted vertex. All the statements corresponding  to merged vertices  constitute a functionality to be extracted out of the native method. Such a functionality is called as \textit{extract method opportunity} (EMO).  After extraction, an EMO is placed into a new method body.

\textit{Application of segmentation:}
In our example, the first control vertex that is analyzed is vertex 19 (if block). The block at 19 has two control dependent statement 20 and 21. Vertex 21 represents a control block at else-statement and consists of one control dependent statement 22. It can be seen that block at 19 has no outgoing data-edge connecting to any vertex outside of the block (i.e. after vertex 22). Thus, block 19 is not producing any result usable by rest  of the code. Hence, it is not considered as an extraction opportunity, and  it is collapsed (contracted) into  vertex 19 as shown in Figure~\ref{fig:contractingBlocks}(a). 

\vspace{0.5mm}

Control block rooted at vertex 16 (if-block) is analyzed next. This block does not define (i.e. assigns value to a variable) any variable. Hence, it is not inspected for extraction. The block is not contracted because it is control dependent on vertex 15. 
Control block rooted at vertex 15 (iterative block) is analyzed next. It can be seen that it has an outgoing data edge from vertex 18. Hence, this block is further examined for assessing its extraction possibility. Lack of computational strength measure for this block happens to be 0.5, that is higher than a default threshold of 0.41. So, this block is not selected for extraction. As this block is not control dependent on any other block,  it is contracted as shown in Figure~\ref{fig:contractingBlocks} (b). Now, an exclusive data dependency on the contracted block can be observed in form of a data edge $\langle14,15\rangle$. Such a dependency is identified by source vertex and such a data dependency between two vertices shows strong association between them. Hence, as the next step, vertex 14 is merged with vertex 15 by contracting edge $\langle14,15\rangle$ as shown in Figure~\ref{fig:contractingBlocks} (c).

Further, LoCS for control block 8 is measured since, the block has one vertex 11, which connects with another vertex outside the block through a data edge. This outgoing connection represents computation of a result at statement 11 used outside the block. Block 8 has two vertices (9, 11) with at least one outgoing data edge and two exclusive source vertices (6, 7). So, LoCS value for the block is 0.25, which is lower than the threshold 0.41. Hence, the block is contracted to a single vertex as shown in Figures~\ref{fig:contractingBlocks} (d) and (e).

Figures~\ref{fig:contractingBlocks} (f)-(h) trace the  subsequent contraction sequence in bottom up manner. The final graph in   Figure~\ref{fig:contractingBlocks}(h) is called the segment graph, which shows  two methods. The colored vertex shows the extracted method, and remaining graph is the base code in which the extracted method can be plugged in as a call. 
The extracted method represented by colored vertex with label \textit{"1-13"} represents block of IR statements from index 1 to 13. We can observe that this block corresponds to source code statements responsible for computing \textit{n}th Fibonacci term.  It gets extracted as a method \textit{FiboPrime3} in Figure~\ref{fig:FiboPrime}(b).

Figure~\ref{fig:FiboPrime} shows the  refactored code as per suggestions produced by the segmentation approach. The segment suggestions as produced by the algorithm can be used in a  computer assisted refactoring process. Segmentation algorithm can also be tuned for a customized extraction based on application context.

\begin{figure*}[!h]   
\begin{minipage}{\linewidth}
\begin{minipage}{.45\linewidth}
\begin{footnotesize}
\textbf{Input Code:}
\begin{lstlisting}[language=C] [basicstyle=\footnotesize]
void FiboPrime()	{
  int i, n, a, b, t;
  printf("Enter value of n (>0)");	
  scanf("%d",&n);
  a = 0;
  if (n ==1)
    printf("Fibo Term is %d ",a);
  else	{
    b = 1;
    for (i = 3; i <=n ; i++){
      t = a + b;
      a = b;
      b = t;
    }
  }
  printf("Fibo Term is %d ",b);
  for (i=2 ; i<=b/2;i++){
    if (b%i == 0)
      break;
  }
  if (b<=1 || i <=b/2)
    printf("Not Prime");
  else
    printf("Prime");
}

\end{lstlisting}
\vspace{1.5cm}
(a) Input source code
\end{footnotesize}
\end{minipage}
\vline
\vline
 \hspace{.6cm}
\begin{minipage}{.4\textwidth}
\begin{footnotesize}
\textbf{Refactored Code:}
\begin{lstlisting}[language=C][basicstyle=\footnotesize]
int FiboPrime3(){
  int i, a, b, n, t;
  scanf("%d",&n);
  a = 0;
  if (n ==1)
    printf("Fibo Term is %d ",a);
  else	{
    b = 1;
    for (i = 3; i <=n ; i++){
      t = a + b;
      a = b;
      b = t;
    }
  }
  printf("Fibo Term is %d ",b);
  return b;
}
void FiboPrime()	{
  int i, b;
  printf("Enter value of n (>0)");	
  b =FiboPrime3();
  for (i=2 ; i<=b/2;i++){
    if (b%i == 0)
	    break;
  }
  if (b<=1 || i <=b/2)
    printf("Not Prime");
  else
    printf("Prime");
}

\end{lstlisting}
(b) Refactored code
\end{footnotesize}
\end{minipage}
\end{minipage}
\captionof{figure}{Fibonacci Prime source code and corresponding refactored code by segmentation}
\label{fig:FiboPrime}
\end{figure*}

\begin{figure}[h]
\centering
\includegraphics[scale = .25]{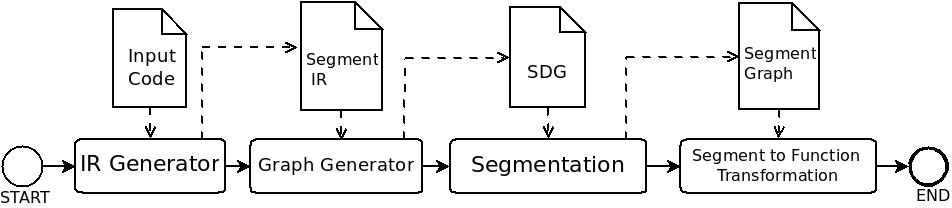}
\caption{Architecture of segmentation based refactoring}
\label{fig:Architecture}
\end{figure}



A high level view of the segmentation approach is shown in BPMN  notation in Figure~\ref{fig:Architecture}. The Figure shows various phases of graph transformations. The first phase transforms the input code to segment IR using an IR generator tool. The segment IR is then used to generate the base SDG, which is used as input to the segmentation algorithm. Segmentation results in a \textit{segment graph}, which provides a solution for refactoring in terms of functions. 


\section{Segmentation}
The segmentation algorithm is now developed in this section in terms of the concept and terms introduced in the previous sections.
\textit{Segmentation} is a transformation of the SDG  into  a \textit{Segment Graph}. Each vertex of the segment graph represents a \textit{segment}, which is a candidate function. The \textit{segmentation} approach exploits both data and control dependency to cluster statements into segments. Segmentation is a process of graph shrinking. 

The SDG is shrunk into a segment graph by repeatedly carrying out three activities enumerated below on each control block vertex in SDG. Each activity shrinks the SDG by means of \textit{contraction} of specific edges. Contraction of an edge results in merging of two vertices, creating a new vertex. The new vertex is assigned the label as that of the segment Id (seg id, Section \ref{Sec:SegId}). As a result,  a bigger subgraph eventually gets reduced into a vertex.
Each of the three activities of segmentation is separately dealt with in  the subsections that follow.

\begin{enumerate}
 \item \textit{Control Edge Contraction:} 
 This activity is carried out on control blocks or inner control blocks in the program.  Every such block is contracted to a single vertex by contracting each of the associated control edges. After contraction, such a block is further investigated for merger through associated data dependencies in next two steps. 
 With each merger the corresponding control block expands and the SDG shrinks.
 
 \item \textit{Exclusive Source Contraction:} It identifies statements providing data exclusively to a control block and merges them into the block.

 \item \textit{Sequential Data Dependence Contraction:} Here, we identify chains associated with block resulted from step 2 (Chains are defined in Section~\ref{Sec:Chain}). Those  chain structures that are functionally coherent with the task implemented by a block are merged into it. 
 
\end{enumerate}

\subsection{Control Edge Contraction (CEC)}\label{ncec}
This section presents a \textit{bottom up} approach, which explores control blocks for contraction in reverse order of their appearance in the code. Such an approach is effective in extracting inner segments from the code containing multilevel nested control blocks. Processing the blocks in reverse order of appearance (bottom-up in source code, and inner first in nesting) makes it possible to first focus on the core (inner blocks) of the functionality,  and then to expand it by including supporting subtasks such as data supplier (e.g. initialization) and chained activities.

The CEC activity aims at identifying and grouping statements (edge contraction) of a control block, which contribute to  a distinct functionality.   
Such a   functionally distinct control block is identified the help of  metrics \textit{lack of computational strength} (LoCS) and \textit{parent affinity} (PA). 
LoCS is applied to a control block to investigate if it contains a distinct functionality.
Once a control block is qualified for extraction using LoCS metric, it is also investigated for merger with its parent control block, using PA. These metrics are defined below.

%
%
%

\subsubsection{Lack of Computational Strength (LoCS):} \label{Sec:LoCS}

Metric LoCS is used to rank control blocks considered by the CEC method. It measures the computational strength of a given control block in terms of the count of producer vertices that directly or indirectly contribute to relay vertices of the block.  It may be noted, a producer vertex represents exchange of data among statements which is a crucial factor in  identifying distinct functionality.

\begin{figure*}[h]   
\begin{minipage}{\linewidth}
\begin{minipage}{.37\linewidth}
\begin{lstlisting}[language=C] [basicstyle=\footnotesize]

int Sum()
{
 int a, sum=0, i;
 for (i=0;i<5;i++)
 {
  scanf("%d",&a);
  sum += a;
 }
 printf("sum =%d\n",sum);
}
\end{lstlisting}
 \vspace{.1cm}
\centerline{(a) Source Code}
\end{minipage}
\vline
\vline
 \hspace{0.5cm}
\begin{minipage}{.25\textwidth}
\vspace{9 mm}
0. assign sum\\
1. assign i\\
2. loop i 3\\
3. \hspace{2mm} input a\\
4. \hspace{2mm} assign sum sum a\\
5. \hspace{2mm}  assign i i\\
6. output sum\\
\\\\
  \vspace{0.2cm}
\centerline{(b) Segment IR}
\end{minipage}
\vline
\vline
\begin{minipage}{0.25\textwidth}
 \includegraphics[scale=0.3]{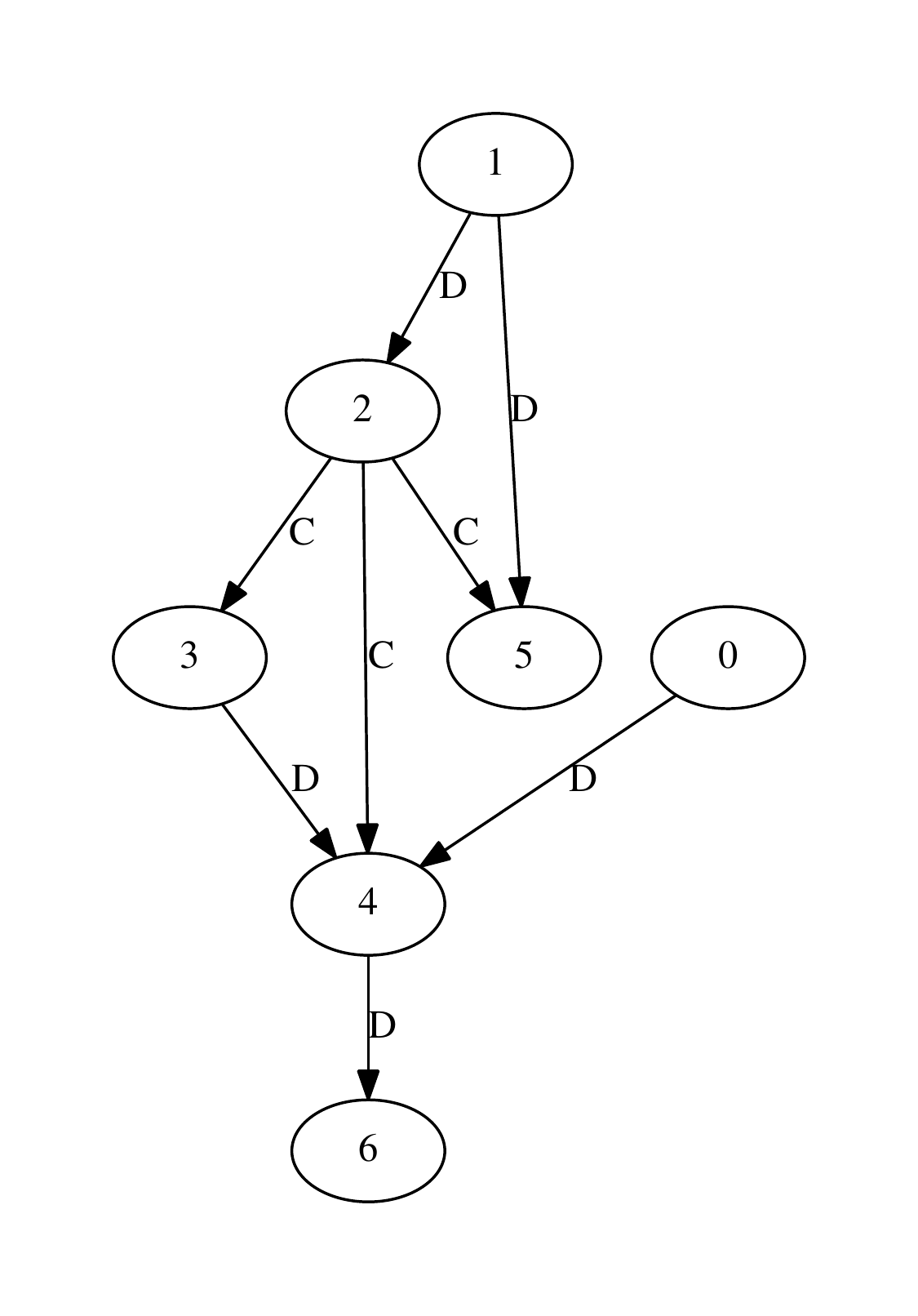}
 \centerline{(c) SDG}
\end{minipage}

\end{minipage}
\captionof{figure}{Non-contributing and Contributing Producers}
\label{fig:loop_index_producer}
\end{figure*}

It can be noted that a producer vertex may not always contribute to a relay vertex, though it may drive a control block. For example,  as shown in the  code given in Figure~\ref{fig:loop_index_producer}, loop variable \textit{i} drives the loop, but is not  used in the computation of the result. Hence, in this example,  producer vertex 1 does not contribute to relay vertex 4. However, vertices 0 and 3 are producers, which contribute to relay vertex 4. The formulation for LoCS is given below.

Let  $b_v$ be a block rooted at a primary control vertex $v$ in SDG $G$. Block $b_v$ is identified by CEC, and is a candidate opportunity.
Let $Relay(b_v,G)$ be a set of all relay vertices in $b_v$. Boolean expression   $Path_D(u,r,b_v,G)$ indicates if there exists a path composed of data edges from vertex $u$ to vertex $r$ in $b_v$. Function $Producer(b_v,G)$ returns the set of all producer vertices in $b_v$. Function  $RelayShare(b_v,r,G)$ provides a set of all producer vertices in $b_v$ from each of which, at least one relay vertex is reachable through a path composed  only of data edges. In other words, $RelayShare(b_v,r,G) = \{u:u \in Producer(b_v,G), \exists r\in Relay(b_v,G) \land Path_D(u,r,b_v,G)\}$.

Even if a producer is not a member of $RelayShare(b_v,G)$, it is a part of computational unit of the block. The set of producer  vertices that are not members of any $RelayShare$ is represented by function $NonRelayShare(b_v,G)$. Count $\#TotalRelayShare$ represents the aggregate count of all the data supplies to  relay vertices.  Data supplies are data edges, not supplier vertices, since every supply is independently counted.  $Relay(b_v,G)$ represents the set of relay statements in $b_v$. Count $\#Relay$ is  the cardinality of set $Relay(b_v,G)$. 

 Lack of Computational Strength (LoCS) is computed  as the ratio of number of relay vertices in a block to the aggregate  computational strength of the block. The denominator is the count of total supplies associated (both internal and external) with the block.
 It is computed as the sum of $\#NonRelayShare$  and   $\#TotalRelayShare$. Also, $\#TotalRelayShare$ represents the count of total sharing from the block to its outer context. Heavy sharing implies higher computational strength, which indicates lower values of LoCS. The same argument applies to $NonRelayShare$.
These quantifications are listed in the box below.

\noindent\fbox{%

\parbox{0.47\textwidth} {
\begin{footnotesize}
$$AllRelayShare(b_v,G) = \bigcup_{r\in Relay(b_v,G)}{RelayShare(b_v,r,G)}$$
$$NonRelayShare(b_v,G) = Procedure(b_v,G) \setminus AllRelayShare(b_v,G)$$ 

$$\#TotalRelayShare(b_v,G) = \sum_{r\in Relay(b_v,G)} {|RelayShare(b_v,r,G)|}$$
$$\textit{\#Relay}  = |Relay(b_v,G)|$$

\begin{equation*}
LoCS(b_v,G) = \frac{  \#Relay} {\#TotalRelayShare + \#NonRelayShare } \\
\end{equation*}

\end{footnotesize}
}
}

\vspace{.2cm}
The lower the value of $LoCS$, the better is the possibility of extracting block $b_v$ as a separate function. A control block is accepted as a segment only if (i) value of $LoCS(b_v,G)$ is less than a threshold (we use 0.41 as threshold which is obtained from experimentation), and (ii) its parent control block (if it exists) contains a significant separate computation. Such computations may share data  with inner control block (i.e. $b_v$) in the form of parameters or return value.

\begin{figure}
\begin{tabular}{p{4cm}p{4cm}}
 \includegraphics[scale=0.17]{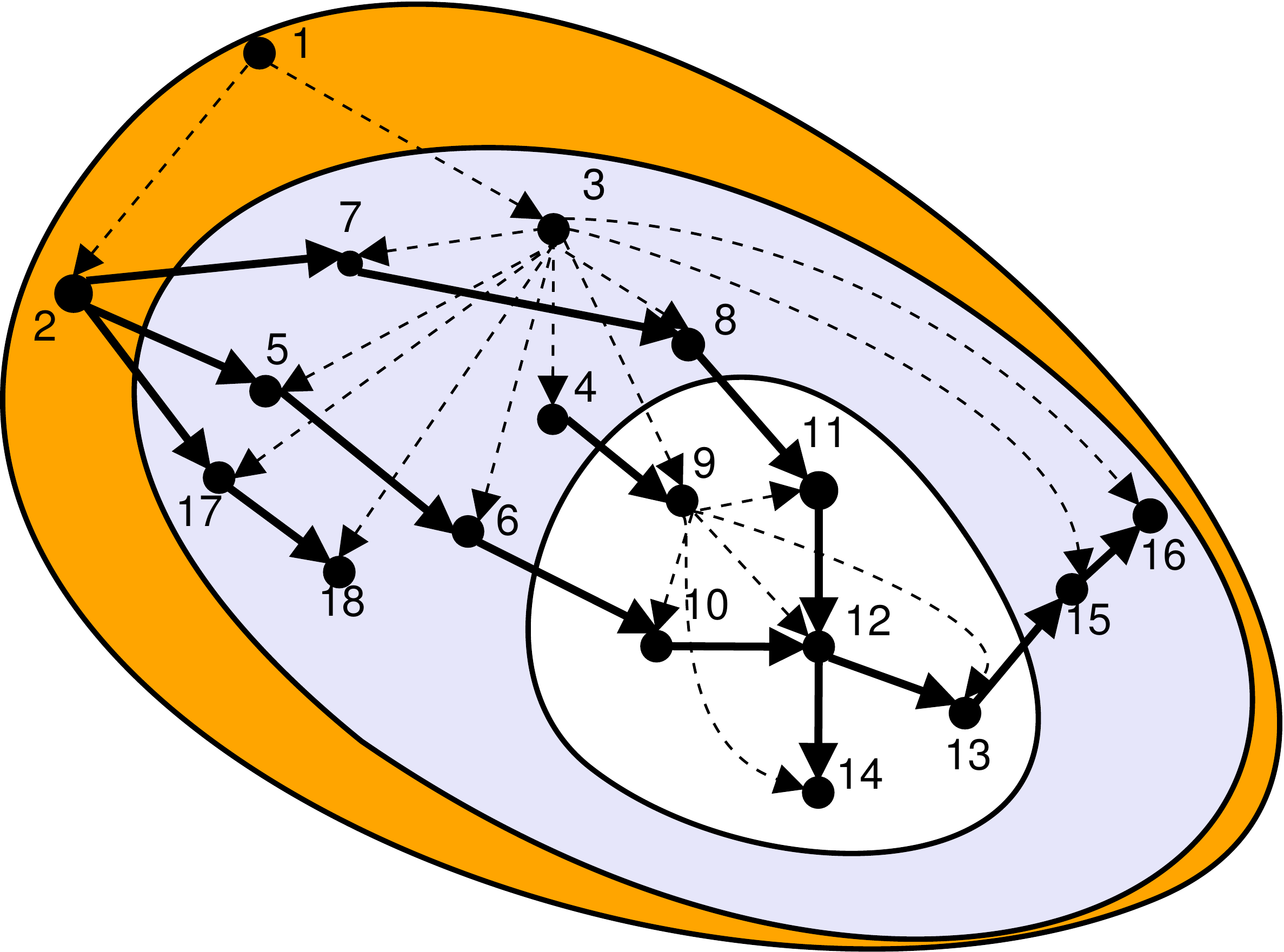} &
 \includegraphics[scale=0.17]{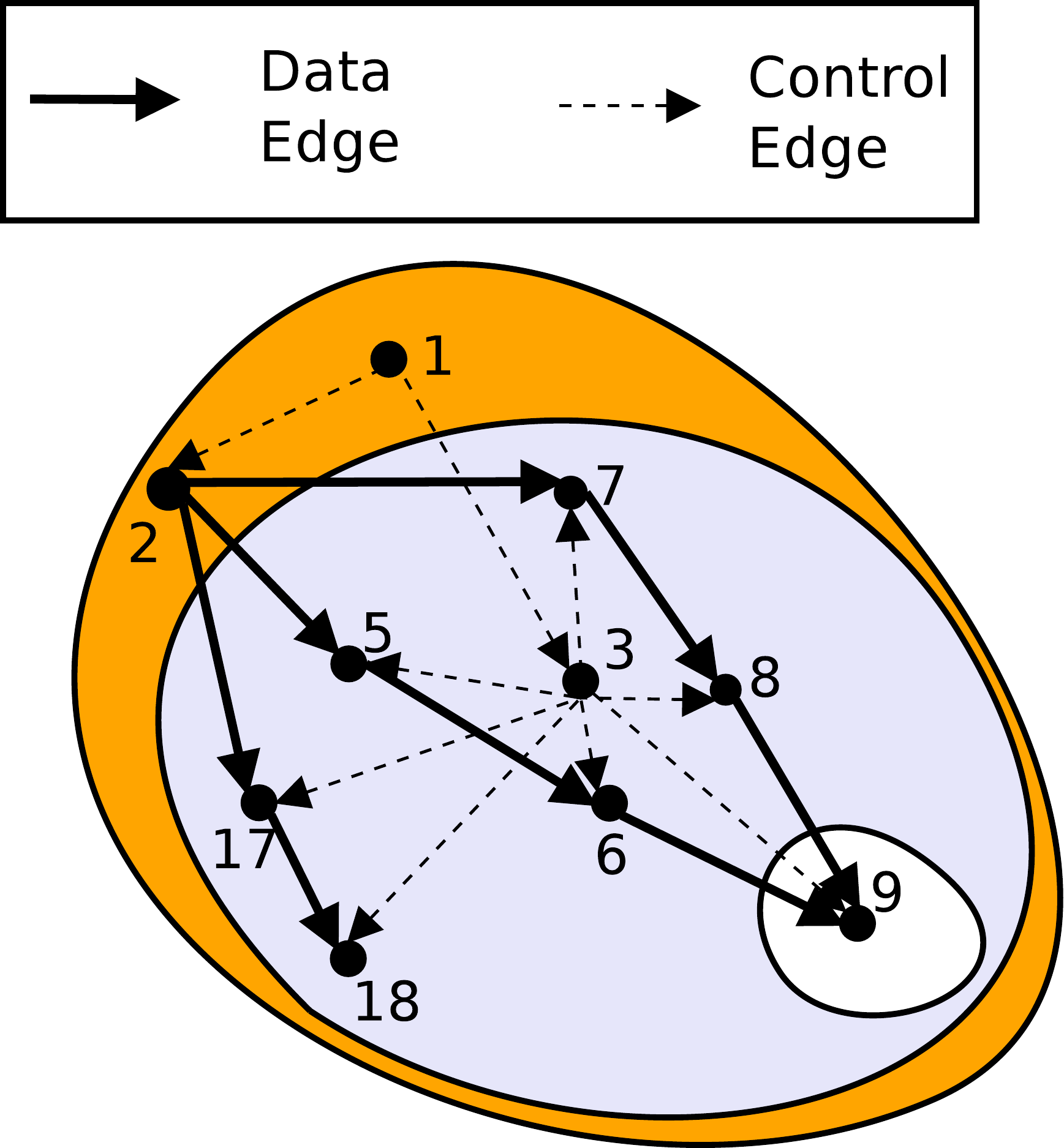} \\ 
 (a) Original SDG & (b) Contraction of block at 9 (V=\{4,9-16\}) \\
 
\end{tabular}

\caption{A sample SDG with nested Control Blocks}
  \label{fig:exampleSDG}
\end{figure}

 \noindent{\em Example Use of LoCS for CEC:} An example SDG is shown in Figure~\ref{fig:exampleSDG}(a). It  contains three nested control blocks. In the figure dotted arrows represent control edges, and thick arrows represent data edges. It can be seen that the outermost block is rooted at vertex 1, middle block is rooted at vertex 3, and the innermost block is rooted at vertex 9. Table~\ref{table:exampleBlock9} lists the attributes associated with control block 9. Since the block has at least one relay statement (vertex 13), it is considered as a candidate for extract method opportunity and its LoCS is computed. As the table shows, block 9 has four vertices with one or more outgoing data edges (10-13), one exclusive source vertex (vertex 4) and one relay vertex (vertex 13). Therefore, LoCS for the block is $\frac{1}{5}$, which is below  0.41, the default threshold used for this example. Hence, the block is contracted, and further, the exclusive source vertex and the outgoing chain (consisting of vertices 13,15,16) are merged to it. Figure~\ref{fig:exampleSDG}(b) shows the SDG after contraction of the block at vertex 9 (consisting of vertices 4 and 9 to 16).
 After contraction of this control block, we need to investigate the parent block at 3 as per the bottom up approach of CEC. The contraction of parent block is explained next.

\begin{table}
\centering
\begin{tabular}{|c|c|c|}

\hline
 \textbf{Attributes} & \textbf{Vertex Set} & \textbf{Count} \\ \hline \hline
 Relay Nodes & \{13\} & 1\\ \hline
 Sink Nodes & \{14\} & 1 \\ \hline
 Exc. Source & \{4\} & 1 \\ \hline
 Producer Nodes & \{4,10,11,12,13\} & 5 \\ \hline
 RelayShare & \{10,11,12\} & 3 \\ \hline
 NonRelayShare & \{4,13\} & 2 \\ \hline
 Incoming Chains & \{$\langle5,6\rangle, \langle6,10\rangle$\} \& \{$\langle7,8\rangle, \langle8,11\rangle$\} & 2 \\ \hline
 Outgoing Chain & \{$\langle13,15\rangle, \langle15,16\rangle$\} & 1\\ \hline

\end{tabular}
\caption{Analysis of Control Block 9 in Figure~\ref{fig:exampleSDG}}
\label{table:exampleBlock9}
 
\end{table}

\subsubsection{Parent Affinity:} \label{sec:PA}
It measures the functional difference between two nested overlapping control blocks, where the inner control block is identified as a candidate opportunity, leading to  its contraction into a single vertex. It is used to determine whether the inner control block should be merged with the parent control block and be extracted together as a single functionality.

For this purpose,  the distinct functionality implemented in the parent block is quantified in terms of distinct computations (computations which are not associated with inner control block via data exchange). Distinct computations are computations which do not supply or consume data to/from inner control block. 
Based on the degree of  coherence  of computations  the inner block with the parent block,  four cases of merger possibility arise. These are enumerated below. It can be observed that in some of them, the  decision is evident due zero overlapping. But in the case of overlapping, we use this metric to decide the  possibility of merger of the inner block into its  parent block. The decisions taken in each of these four cases are as follows:
(i) Parent block has a distinct relay statement, and all its computations are connected with inner control block: In this case they can be merged into one block due to its close association. (ii) Parent block has a distinct relay statement, and not all or none of  its computations are connected with inner control block: In this case, their merger possibility is decided by PA.(iii) Parent block has no distinct relay statement,
 and it is either empty or all its computations are connected with the inner block: They can be merged into one block due to its close association. (iv) Parent block has no distinct relay statement,
 and none or all its computations are connected with inner block is present: Their merger possibility is decided by Parent Affinity (PA).
The metric \textit{parent affinity} (PA) is defined as follows.

\begin{equation*}
PA(v,G) = 1- \frac{ IndependentNodes(b_p,b_v,G)}{ ParentDataNodes(b_p,b_v, G) }\\
\end{equation*}

where,  (i) $b_v$ and $b_p$ represent the control blocks  respectively at an inner control vertex $v$  and its parent vertex $p$, (ii) \textit{ParentDataNodes} ($b_p,b_v, G$) is the count of  those vertices in $b_p$ from which there is  at least one outgoing data edge. 
(iii) \textit{IndependentNodes} ($b_p,b_v, G$) is the count of those vertices in $b_p$ which are not directly connected with $b_v$ by a data edge (after contraction of $b_v$). 

\noindent{\em Example Use of Parent Affinity for CEC:}
Figure~\ref{fig:exampleSDG} (b) shows the SDG after contraction of the inner block rooted at vertex 9. After the contraction of this block, its  parent block rooted at vertex 3 still contains six vertices in addition to the contracted vertex 9 and vertex 3 itself. Two of these vertices (6 and 8) are directly connected with a data edge to block 9, whereas, vertices 5 and 7 are indirectly connected to block 9 via data edges. In block 3, only two vertices (17 and 18) that do not supply/consume data to/from block 9 remain. Thus, $Producer(b_p,G)$ is the vertex set V = \{5,6,7,8,17\}. Out of these, producer vertices  5, 7, and 17 are not directly connected with inner block 9. They form the set \textit{IndependentNodes}. So,  parent affinity (PA) for block 3 is (1 - (3/5)) = 0.4. Since it is higher than default threshold of 0.34, this  parent block is not merged with inner block 9. Thus, only  inner block 9 with vertices 4, 9-16 is extracted as a separate method.

\subsubsection{Contraction:} Once a control block is qualified for extraction, all control edges originating from it are contracted, resulting in reduction of the subgraph to the control block  vertex  called \textit{target}. The target represents a newly contracted control block constituting a distinct functionality. Remaining two activities of the segmentation investigate  possibilities of merging direct contributing data dependencies into the target.  
Procedure \ref{Alg:NCEC} shows the identification and contraction of control blocks as extract method opportunity. Procedure  $EdgeContraction(v,u,v,G') $ contracts the edge between vertices $u$ and $v$. Vertex resulting from the contraction is given the label $v$ (third argument).
 
\begin{myalgo}
 \KwData{SDG G}

$G'  \leftarrow G $\\
$ctrList \leftarrow GetPrimaryControlVList(G')$ \\
$ v \leftarrow GetLast(ctrList)$\\
\While {$ HasElement(ctrList) $}
{
    $Del(v,List)$\\
    $ count \leftarrow $ \textit{GetRelayCount}$(v,G')>0 $\\
    \If { $count >0$ And $CA(v,G')< THRESHOLD$}
      {
	$v \leftarrow $ \textit{GSI}$(v,G', List)$\\
	$G' \leftarrow  \textit{SDDC} (v,\textit{ESC}  (v,\textit{CCB} (v,G')$))\\
      }
     $v \leftarrow GetLast(List)$\\
   
 }
  \Return{G'}
\caption{ Control Edge Contraction (CEC)}
\label{Alg:NCEC}
\end{myalgo}

\begin{myalgo}
 \KwData{Vertex v, SDG G, PCL List}

$G'  \leftarrow G $\\
$p \leftarrow getControlParent(v,G')$

\While {$IsPrimaryCtrlVtx(p)$ And $PA(p,v,G')<PThreshold$}
{
  $Del(v,List)$\\
  $v \leftarrow p$\\
  $p \leftarrow getControlParent(v,G')$
     
 }
  $Del(v,List)$\\
  \Return{$v$}
\caption{Get Segment Id (GSI)}
\label{Alg:GSI}
\end{myalgo}

\begin{myalgo}
 \KwData{SDG G, Vertex v}

$G'  \leftarrow G $\\

\While {$ HasControlEdge(v,G') $}
{
    $ \langle v,u \rangle \leftarrow GetNextCtrlEdge(v,G')$\\
    $ G' \leftarrow EdgeContraction(v,u,v,G') $\\
       
 }
  \Return{G'}
\caption{Contract Control Block (CCB)}
\label{Alg:CCB}
\end{myalgo}

%
%
%

  \subsection{Exclusive Source Contraction} Once a control block (\textit{target}) is contracted, the next activity involves merging of contractible initializations and  input statements, which are those present in the same control region as that of the target.    
  Each of the contractible input/initialization statements provides data to exactly one vertex in the graph produced by  control edge contraction. Such statements in the SDG can hence be identified  by tracking the  \textit{source} vertices with a singular outgoing data edge. 
  Thus, an edge $\langle u,v \rangle $ originating from a source vertex  $u$ is contracted only if its outdegree is one. The new vertex created after contraction of the edge $\langle u,v\rangle$  is assigned the label $v$, thus merging loosely hanging source vertices with the main component to which they are data \textit{sources}.

  It is possible that the newly created vertex is also a \textit{source vertex} for some other vertex. In such cases, the same process is repeatedly applied till the data source chain merges into the target. The merger of the source chain into the target results in a vertex with label of the  target. 
  Figure~\ref{fig:contractingBlocks}(g) shows the output of \textit{exclusive source contraction} applied to the output of control edge contraction.

\begin{myalgo}[b]
 \KwData{ $G'$: The graph produced by CEC, $target$ : newly contracted control block Id }
    $G'' \leftarrow G'$\\
    SourceList $\leftarrow$ \textit{SourceAt}($G''$)\\
    
   \While { SourceList $\neq \phi$ }
    {     
      
        $u \leftarrow $ Next(SourceList) \\
	\If{$Outdegree(u) = $ 1 And \textit{ControlRegionDiff}$(target,u) = $ 0 }
	 {
	    
	    $w \leftarrow Adj(u,G'')$ \\
	    $G'' \leftarrow $\textit{EdgeContraction}($ \langle u,w\rangle,w$,G'')\\
	  SourceList $\leftarrow $\textit{Source}($G''$)\\   
	  }
	  \Else 
	  {
	    SourceList $\leftarrow$ SourceList $-$ $u$\\
	  }
     }
    \Return{$G''$}  
   \caption{ Exclusive Source Contraction (ESC)}
    \label{Alg:Source}
    \end{myalgo}



Procedure~\ref{Alg:Source} outlines the process of exclusive source contraction. The procedure uses 
procedure \textit{Source($G$)}, which  selects and returns the set of all source vertices in graph $G$. Procedure \textit{EdgeContraction()} is described in the previous section.

\subsection{Sequential Data Dependence  Contraction (SDDC)}\label{sec:SDDC} This step groups statements involved in sequential data dependence. For example, consider k statements $s_0, s_1, ..., s_{(k-1)}$ in a program such that each statement $s_i$ is data dependent on statement $s_{i-1}$ where $1 \leq i \leq (k-1)$. Such statements can be identified in the SDG by tracking \emph{chains} of data dependency. These statements are grouped by contracting edges present in \textit{chains}.
A chain structure may cross beyond the control region of the target block. So, a truncation of the chain is  required. Such a chain is truncated at the border edge connecting two of its vertices  present in different control regions. Thus, only those vertices of the chain that are in the same control region are considered for merger.
 A chain can be associated with the target in two ways, as incoming or as outgoing. It can be observed that an incoming chain into the target supplies data to the block, whereas, chain outgoing from the target consumes data produced by the block. 
 Each chain structure represents a subtask. But the challenge lies in identification of  chain structures that keep the resulting block functionally coherent.  The selection criteria for chain structures is given below.

\subsubsection{Chain Structure Selection:}
  If there is only one chain incoming to the target vertex, it is merged into the target block, which is  extracted as a method. The same applies to the lone outgoing chain from the target vertex. Such chain structures show strong association of data dependence with target block.  
  Presence of more than one incoming or outgoing chains can be of two types based on size. These cases are handled as follows:
  \begin{enumerate}
   \item \textit{Unit chains:} 
    These chains are of length one. If incoming, such a chain represents an assignment of  a variable or an input call such as scanf().    
    If outgoing, it represents a return statement, or an output statement or a statement which consumes but does produce a data value as shown in edge $\langle4,6\rangle$ in Figure~\ref{fig:loop_index_producer} representing use of variable \textit{sum} in a printf statement outside the control block corresponding to the for loop.

  All such chains can be merged to the target block. An added benefit of  merger of incoming unit chain is that the definitions (of variables) required for the functionality in the block will be at one place. If the chain head has two or more predecessor then merger of the unit chain increases the parameter list of the extracted method. 
  Similarly, an outgoing chain that is  of length one and which does  not extend outside of control region of target block, either represents an output statement or a  data  sink statement.   A sink statement uses the result but does not create a data dependency on any succeeding statement. 
An added benefit of  merger of outgoing unit chain is that it  avoids extra efforts in managing more than one return arguments.
  
%
%

  \item \textit{Longer chains:}   These chains are of length more than one.
  All such chains represent subtasks. In this case, after merging all one length chains if only one long chain remains,  it is merged with the target block. However, if there are more than one long chains associated with the block, user intervention may be required to rank functional closeness of the chains with  the target block. Hence, at present, segmentation  does not   automatically  merge such chains with the target block. 
  \end{enumerate} 
 
Figure~\ref{fig:contractingBlocks}(h) shows the output of \textit{sequential data dependence contraction}  applied on  graph produced by exclusive source contraction in Figure~\ref{fig:contractingBlocks}(g). Sequential data dependence contraction is implemented in Procedure \ref{Alg:Chain}.

 \begin{myalgo}
 \KwData{$ G''$: The graph produced by SESC, $v$: Newly contracted Block }
    $G''' \leftarrow G''$\\
    incoming $\leftarrow$ \textit{ChainsAt}($v,G''$)\\
    outgoing $ \leftarrow$ \textit{ChainFrom}($v,G''$)\\
    count $\leftarrow 0$ \\
   \While { incoming $\neq \phi$ }
    {     
      
        $c \leftarrow $ Next(incoming) \\
        \If {length(c)==1}{ 
	  \textit{EdgeContraction}($ \langle u,v \rangle,v,G'''$)
        }
        \Else {
	  count $\leftarrow $ count + 1\\
	  tmpchain $\leftarrow$ c
        }
    $incoming \leftarrow incoming$  $-$ $c$
    }
    \If{count == 1}{
      \ForEach  { $ \langle u,w \rangle \in c$ }
      {
	  \textit{EdgeContraction}($ \langle u,w \rangle,u,G'''$)
      }
        \textit{EdgeContraction}($ \langle u,v \rangle,v,G'''$)
    }
    \If{length(outgoing)==1}{
      $c \leftarrow $ Next(outgoing) \\
      \ForEach  { $ \langle u,w \rangle \in c $ }
      {
	  \textit{EdgeContraction}($ \langle u,w \rangle,u,G'''$)
      }
      \textit{EdgeContraction}($ \langle u,v \rangle,v,G'''$)
    }
    \Return{$G'''$}  
   \caption{Sequential Data Dependence Contraction (SDDC)}
    \label{Alg:Chain}
    \end{myalgo}

\subsection{The Segmentation Algorithm}
The complete algorithm is listed as Algorithm \ref{Alg:SegSlice} in the form of nested function calls nesting the contraction operations in segmentation described above. The input is SDG $G$, and the output is segment graph $S$.

\begin{algorithm}
 \KwData{SDG G} 
  \Return (CEC(G))
\caption{Segmentation}
\label{Alg:SegSlice}
\end{algorithm}


\section{Evaluation}

As discussed previously, the proposed approach aims at suggesting minimal set of extract method opportunities while maximizing  functionally relevant suggestions. Table~\ref{table:OSS_stat} lists the chosen sources for the evaluation case studies, mentioning the sizes of the programs and opportunities pre-marked by  developers.

\begin{table}[h]
  \centering
    \caption{Evaluation Case Studies}
   \begin{tabular}{|c|p{2.5cm}|p{3cm}|}
   \hline
   \textbf{Case Studies}& \textbf{\#Methods (Program Size)} & \textbf{\#Extract Method opportunities}  \\ \hline \hline
   Segmentation  &3 & 18\\ \hline
   JUnit  \cite{silva2014recommending} & 25 & 25 \\ \hline
  JHotDraw  \cite{silva2014recommending} & 56 & 56 \\  \hline
   XData \cite{bhangdiya2015xda} &  32 & 110  \\ \hline
		
      \end{tabular} 
   
  \label{table:OSS_stat}
  \end{table}

The evaluation of the approach consists of two parts. In the first part, we discuss the process of tuning segmentation for finding promising threshold values on one program in C that was manually mapped to  intermediate representation. 
The program chosen was a non-refactored synthetically created version of our own implementation of the \textit{segmentation} algorithm in this paper. The results obtained matched with manually refactored version. These results are discussed later in this section.

In the second part, we apply a tool based on the proposed technique on three open source Java programs and present the results, comparing them with another approach taken by the JDeodorant  \cite{tsantalis2011identification} tool.
The results obtained after refactoring and also a feedback obtained from a developer of one of the programs are discussed.
Two of the three chosen Java-based OSS case studies, \textit{JUnit} and \textit{JHotDraw}, were created by silva et al.  \cite{silva2014recommending} by inlining the callee methods inside the caller method.  Both have been used earlier in the literature  \cite{silva2014recommending}, 
 \cite{charalampidou2017identifying} as benchmarks for long method extraction. The third OSS project chosen is the \textit{XData} grading system for queries  \cite{bhangdiya2015xda}, which contains method sizes of sufficiently varied lengths (30-200 LOC) for our experimentation.

  \subsection{Performance metrics used for comparison}

  To measure the performance we use precision, recall and F measure metrics. Extract method opportunities listed by the our approach and JDeodorant tool are compared with pre-marked opportunities, and the shares of True Positives, False Positives,  and False Negatives are identified as mentioned below.

\vspace{0.15cm}
\noindent\fbox{%

\parbox{0.5\textwidth-0.5cm} {
 \begin{itemize}
  \item \textit{True Positive (TP):} opportunity identified by both the developer and the algorithm
  \item \textit{False Positive (FP):}  opportunity identified only by the algorithm
  \item \textit{False Negative (FN):} opportunity identified only by the developer 
       \end{itemize}

 $Precision = \frac{\#TPs}{Retrieved\; opportunities} = \frac{\#TPs }{ (\#TPs + \#FPs)}$\\
 
 $Recall = \frac{\#TPs}{Relevant \;opportunities} = \frac{\#TPs }{ (\#TPs + \#FNs)} $\\
 
 $F measure = \frac{2*(Precision * Recall)}{(Precision + Recall)}$\\

    } 
 }
 \\
 
 \noindent
 {\em Match Tolerance:} We have analyzed the effect of applying \textit{match tolerance} values of 1 to 3 statements on the identification of opportunities by our algorithm. A deviation of a match within the tolerance range in both
 direction is acceptable.
 The same tolerance value is used for the JDeodorant tool for the purpose of comparison.

\subsection{The Process of  Tuning}
 The proposed segmentation approach primarily uses three parameters to control the number of extract method opportunities (suggestions), namely, \textit{LoCS}, \textit{PA}  as discussed earlier,
 and a flag \textit{NoRelayExtract}. This flag is used as follows. By default, segmentation computes LoCS metric for  blocks containing at least one variable (relay vertex), which is assigned inside and is used outside the block. However,  in some cases, a method may not contain a relay vertex, but may still represent a distinct functionality. To account for such methods, this flag can be set to value \textit{True}, which allows the segmentation algorithm to compute values of LoCS for such blocks also. This is achieved by setting the numerator value $\#Relay$ to one in equation ~\ref{Sec:LoCS}.
  Blocks for which LoCS is not computed are not considered into output suggestions.  
 To find the appropriate threshold values for aforementioned parameters, we performed a study on an older version of segmentation approach and analyzed the extract method suggestions against varying thresholds for LoCS and PA. The values of LoCS and PA found in the  case study discussed in this subsection are used for three other case studies discussed in the next section.
 Now, we discuss the method of tuning LoCS and PA.

 An older \textit{ modularized base code} of segmentation implemented in C was first transformed into \textit{non-modularized  code},  by \textit{unfolding} a few functionalities into their caller functions. New functions of varying sizes were formed by unfolding  functions in the source.  The non-modularized code for our evaluation that is generated from  the above method consisted  of a total of 28 distinct functionalities spread over 13 functions. Ten out of these 13 functions contain a single distinct functionality, resulting in three large functions containing 18 functionalities all in all. 
The original modularized base code was considered for comparing the results.
Table~\ref{table:PrecisionSegmentation} shows the precision and recall values for the transformed  code. It is observed that  threshold values of 0.41 and 0.34  respectively for LoCS and PA, provided the best results for this case study. Flag `NoRelayExtract' was set to \textit{False}.
These parameter settings for segmentation are used as  default seed  values/setting for next three applications, which are  tuned as per the need of the application.

 \begin{table}[t]
  \centering
   \caption{Precision and Recall obtained after tuning}
   \begin{tabular}{|p{3cm}|c|p{0.12cm}p{0.12cm}p{0.23cm}|p{.8cm}p{.8cm}|}
   \hline
   {\footnotesize Methods to be refactored}& {\footnotesize \#LoC} &{\footnotesize TP} & {\footnotesize FP} & {\footnotesize FN} & {\footnotesize Precision} & {\footnotesize Recall} \\ \hline
   {\footnotesize \#1 FindExcSourceNodes}& 55& 3 &1 & 2 &0.75 &0.60\\ \hline
   {\footnotesize \#2 InitializeSourceList} & 23 & 2 &1  &0 &0.66 &1.00\\ \hline
   {\footnotesize \#3 FindChain}& 137& 5 & 3& 6& 0.62&  0.54\\ \hline
   \end{tabular}
 
  \label{table:PrecisionSegmentation}
  \end{table}

\subsection{Application to OSS code}
JUnit and JHotDraw projects provide already marked extract method opportunities, whereas for XData grading system, we requested one of the developers to manually identify the extract method opportunities. The developer in this case  was a  research scholar with  some  industry experience. A brief description about our approach was provided to the developer before marking.
Table~\ref{table:tune_seg} shows the settings for which we obtained satisfactory results for all three studies.  We find that segmentation with default setting is not able to extract any extract method opportunity for JUnit, whereas in case of JHotDraw,  it provides better precision than JDeodorant with a low recall.  
 \begin{table}[b]
  \centering
  \caption{Tuning segmentation for best results }
   \begin{tabular}{|c|c|c|c|c|}
   \hline
   \textbf{Projects}& \textbf{LoCS value} & \textbf{PA value} & \textbf{'NoRelayExtract'}  \\ \hline

   JUnit & 0.41 & 0.34 & \textit{True} \\ \hline
  JHotDraw & 0.41 & 0.34 & \textit{True}\\  \hline
   XData&  0.41 & 0.34 & \textit{False} \\ \hline
		
      \end{tabular} 
   \label{table:tune_seg}
  \end{table}

 \vspace{0.2cm}

Table~\ref{table:SuggestionCount} shows the number of opportunities suggested  by both approaches for each study. As it can be observed that segmentation and  JDeodorant are comparable except for XData system, for which the latter suggested far more opportunities than marked by the developer.

 \begin{table}[b]
  \centering
  \caption{Extract method opportunities for all methods}
   \begin{tabular}{|c|p{1cm}|p{1.2cm}|p{1cm}|p{1cm}|}
   \hline
   \textbf{Tools/OSS}& \textbf{JUnit (25)} & \textbf{JHotDraw (56)} & \textbf{XData (110)} & \textbf{Total (191)}  \\ \hline
		
   Segmentation &11 & 72&117&200\\ \hline

   JDeodorant 	& 14 & 63 &202 &279\\ \hline
		
      \end{tabular} 
   
  \label{table:SuggestionCount}
  \end{table}

  Tables ~\ref{table:PrecisionJHotDraw}, \ref{table:PrecisionXData}, and  \ref{table:PrecisionJunit} present a detailed comparison of our approach with JDeodorant  over the three applications. The experiments were conducted for three \textit{match tolerance }values of 1,2 and 3.
   It can be observed that JDeodorant performed better than segmentation for JUnit, whereas in case of JHotDraw and XData, segmentation performed  better. The results are now analyzed with pointers to future work.


\begin{itemize}
 \item \textit{Accuracy in suggestions:} 
    Segmentation outperforms JDeodorant in precision as well as recall except in case of JUnit as shown in Table~\ref{table:PrecisionJHotDraw}, \ref{table:PrecisionXData} and \ref{table:PrecisionJunit}. We observed that segmentation has suggested  a few marked opportunities in JUnit, but they were wrapped with an additional associated \textit{if-else} block or a \textit{try-catch} block. Such cases resulted in lower precision.
    In future, these features can be explored as alternate suggestions or improvements,  assisting the method of manual refactoring. 
    
  \item \textit{ Effect of NoRelayExtract Flag:}
   This flag can be set to \textit{True} if the opportunities of interest do not have a relay vertex in the SDG. It can be seen from Tables~\ref{table:PrecisionJHotDraw} and \ref{table:PrecisionJunit} that in the case of JHotDraw and JUnit, the performance of the segmentation approach improved when the flag was set to \textit{True}. The same is not observed for XData, in which, there are large functionalities. In this case the flag set to \textit{False} provides better overall performance as shown in Table~\ref{table:PrecisionXData}. In future this selection can be incorporated through an application specific decision making.
   
  \item \textit{Performance over long methods :}
  To measure the performance of both approaches over long methods, we selected methods with more than 150 LOC. Table~\ref{table:Longmethods} tabulates the results for the segmentation approach. JDeodorant did not suggested any matching suggestions in these cases. Thus, segmentation provided significantly better performance over JDeodorant in terms of precision and recall. To note, the recall crossed over 50.
  We haven't  observed the performance of the approach over very large method such as  in range 200+ LOC. In such cases although the programmer may a set of functionality as one unit, the tool may suggest a break-up. The approach can be assessed for identification of such large units to assist in refactoring.

  \item \textit{Comparing with expert's markings:}
      One of the primary motivation behind choosing XData system was to apply the approach over non-synthetic long method, and also to compare the result with expert's markings. We note that both the tools have identified and suggested a  similar set of opportunities that were forming blocks bigger than marked opportunities crossing  the marking tolerance. Hence they were not classified as matches. However, we did not perform further
      analysis of scrutinizing the results of manual markings. This line can be further investigated for using the tool in assistance with the expert or to evaluate manual decisions.

    \item {The effect of programming language features:}
    The proposed algorithm   identified  majority  opportunities present in the considered applications compared to JDeodorant. However, we noticed that a few of the opportunities that were not identified  were Java specific, especially anonymous inner classes. Inclusion of support for language specific plugins to the SDG model may further improve the performance of the approach. 
    
%
%
%
%
%
%
%
\end{itemize}

  \begin{table}
  \centering
  \caption{Comparision for JHotDraw}
   \begin{tabular}{|p{2.7cm}|p{1cm}|p{1cm}|p{.7cm}|p{1.3cm}|}
   \hline
   \textbf{Tools}& \textbf{Tolerance} & \textbf{Precision} & \textbf{Recall} & \textbf{F measure}  \\ \hline

				& 1 & 15.27 & 19.64 & 17.18\\ 
   Segmentation 		& 2 & 23.61&30.35 & 26.56\\
    (`NoRelayExtract'=\textit{True})	& 3 & 30.55& 39.28 & 34.37\\ \hline

				& 1 & 7.83 & 8.92 &8.40 \\ 
   JDeodorant 			& 2 & 11.11 &12.5& 11.76\\
				& 3 & 20.63& 23.21 & 21.84\\ \hline \hline
				
				& 1 & 12 & 5.35&7.4\\ 
   Segmentation 		& 2 & 20&8.92&12.34\\
  (`NoRelayExtract'=\textit{False})		& 3 & 28& 12.5&17.28\\ \hline
  \end{tabular} 
  \label{table:PrecisionJHotDraw}
  \end{table}
 
   \begin{table}
  \centering
  \caption{Comparision for XData grading system}
   \begin{tabular}{|p{2.7cm}|p{1cm}|p{1cm}|p{.7cm}|p{1.3cm}|}
   \hline
   \textbf{Tools}& \textbf{Tolerance} & \textbf{Precision} & \textbf{Recall}  & \textbf{F measure} \\ \hline
		& 1 & 51.28 & 54.54 & 52.86\\ 
   Segmentation & 2 &  55.55&59.09 & 57.26\\
(`NoRelayExtract'=\textit{False})	& 3 &  57.26& 60.09&59.03\\ \hline
		& 1 & 2 & 3.23 &2.53\\ 
   JDeodorant 	& 2 & 3.46 &6.36&4.48\\
		& 3 &  3.46& 6.36 &4.48\\ \hline
      \end{tabular}

  \label{table:PrecisionXData}
  \end{table}
  
   \begin{table}
  \centering
  \caption{Comparision for JUnit}
   \begin{tabular}{|p{2.7cm}|p{1cm}|p{1cm}|p{.7cm}|p{1.3cm}|}
   \hline
   \textbf{Tools}& \textbf{Tolerance} & \textbf{Precision} & \textbf{Recall} & \textbf{F measure}  \\ \hline
		& 1 & 9.09 & 4 &5.55\\ 
   Segmentation & 2 &  9.09&4&5.55\\
(`NoRelayExtract'=\textit{True})	& 3 &  18.18&  8&11.11\\ \hline
		& 1 & 35.71 & 20 &25.64\\ 
   JDeodorant 	& 2 & 42.85 &24&30.76\\
		& 3 &  42.85& 24 &30.76\\ \hline
      \end{tabular}

  \label{table:PrecisionJunit}
  \end{table}

   \begin{table}
  \centering
  \caption{Performance over long methods (LOC$>$150)}
   \begin{tabular}{|p{2.7cm}|p{1cm}|p{1cm}|p{.7cm}|p{1.3cm}|}
   \hline
   \textbf{Tools}& \textbf{Tolerance} & \textbf{Precision} & \textbf{Recall}  & \textbf{F measure} \\ \hline
		& 1 & 38.29 & 50 & 43.37\\ 
   Segmentation & 2 &  40.42&52.77 & 45.78\\
(`NoRelayExtract'=\textit{False})& 3 &  40.42& 52.77&45.78\\ \hline
      \end{tabular} 
   \label{table:Longmethods}
  \end{table}   
\section{Conclusions}
\textit{Segmentation}, a novel graph-based approach for extract method refactoring was presented. The segmentation algorithm uses a language independent intermediate representation called {\em segment IR}. A {\em structure dependence graph} (SDG) is extracted from the segment IR. The algorithm transforms this graph into a segment graph, which is obtained by contracting subgraphs through {\em edge contraction}. The approach was evaluated on code from three different open source software projects, and promising results were obtained. The approach was found to have worked well on long methods. 
The SDG based framework  developed in the paper is open to  tuning. The approach can be used as a \textit{refactoring  assistant} to aid human expert in identifying modular functionalities through extract method opportunities. Further work in this area includes extension to cover inter-procedural analysis, language specific features such as scoping  and merger of segments across procedures. Impact of language specific features on modularization can also be investigated.
\\
\bibliographystyle{IEEEtran}
\bibliography{SegSlice}
\end{document}